\documentclass[a4paper]{article}

\usepackage[sort&compress,numbers,square]{natbib}
\usepackage[margin=1in]{geometry} 
\usepackage{orcidlink}

\usepackage[figuresright]{rotating}
\usepackage{listings}
\usepackage{pgfplots}
\pgfplotsset{compat=1.8}
\usepgfplotslibrary{statistics}
\usepackage{tcolorbox}
\usepackage{graphicx}
\usepackage{subfigure}
\usepackage{multirow}
\usepackage{lscape}
\usepackage{xcolor}

\usepackage{color, colortbl}
\definecolor{Gray}{gray}{0.9}
\newcolumntype{g}{>{\columncolor{Gray}}l}

\newcommand{\avmo}{{\fontfamily{cmss}\selectfont \mbox{AVMo}}}
\newcommand{\avmc}{{\fontfamily{cmss}\selectfont \mbox{AVMc}}}
\newcommand{\avmr}{{\fontfamily{cmss}\selectfont \mbox{AVMr}}}
\newcommand{\avmrc}{{\fontfamily{cmss}\selectfont \mbox{AVMrc}}}
\newcommand{\rs}{{\fontfamily{cmss}\selectfont \mbox{RS}}}
\newcommand{\umltocsp}{{\fontfamily{cmss}\selectfont \mbox{UMLtoCSP}}}
\newcommand{\pledge}{{\fontfamily{cmss}\selectfont \mbox{PLEDGE}}}

\newcommand{\gcs}{{\fontfamily{bch}\selectfont \mbox{GCS}}}
\newcommand{\eur}{{\fontfamily{bch}\selectfont \mbox{EUR}}}
\newcommand{\rnl}{{\fontfamily{bch}\selectfont \mbox{RnL}}}
\newcommand{\sm}{{\fontfamily{bch}\selectfont \mbox{SM}}}
\newcommand{\sbs}{{\fontfamily{bch}\selectfont \mbox{SBS}}}
\newcommand{\srs}{{\fontfamily{bch}\selectfont \mbox{SRS}}}

\title{Efficient Test Data Generation for MC/DC with OCL and Search}

\author{Hassan Sartaj\orcidlink{0000-0001-5212-9787}$^{1}$, Muhammad Zohaib Iqbal$^{2}$, Atif Aftab Ahmed Jilani$^{1}$, and \\Muhammad Uzair Khan$^{1}$}

\date{
    $^1$National University of Computer and Emerging Sciences, Islamabad, Pakistan \\
    $^2$Quest Lab, Islamabad, Pakistan \\[2ex]%
 }
 
\begin{document}
\maketitle
 
	\begin{abstract}
		System-level testing of avionics software systems requires compliance with different international safety standards such as DO-178C. 
		An important consideration of the avionics industry is automated test data generation according to the criteria suggested by safety standards. 
		One of the recommended criteria by DO-178C is the modified condition/decision coverage (MC/DC) criterion. 
		The current model-based test data generation approaches use constraints written in Object Constraint Language (OCL), and apply search techniques to generate test data. 
		These approaches either do not support MC/DC criterion or suffer from performance issues while generating test data for large-scale avionics systems.			
		In this paper, we propose an effective way to automate MC/DC test data generation during model-based testing.
		We develop a strategy that utilizes case-based reasoning (CBR) and range reduction heuristics designed to solve MC/DC-tailored OCL constraints.
        \textcolor{black}{We performed an empirical study to compare our proposed strategy for MC/DC test data generation using CBR, range reduction, both CBR and range reduction, with an original search algorithm, and random search. }
        \textcolor{black}{We also empirically compared our strategy with existing constraint-solving approaches.}
		The results show that both CBR and range reduction for MC/DC test data generation outperform the baseline approach. 
		\textcolor{black}{Moreover, the combination of both CBR and range reduction for MC/DC test data generation is an effective approach compared to existing constraint solvers.}  
	\end{abstract}
 

	\section{Introduction}
	
	Aircraft avionics are safety-critical systems that are required to comply with international safety standards.  
	DO-178C~\cite{do178} is the prevailing standard for the avionics of aircraft systems.  
	A large part of DO-178C (around 70\%) focuses on verification and validation activities.  
	Testing forms a significant portion of the overall activities required to comply with the standard. 
	According to DO-178C, an important consideration for the highest critical level avionics applications is to test these applications based on the Modified Condition/Decision Coverage (MC/DC) criterion. 
	The MC/DC criterion subsumes different coverage criteria, e.g., statement coverage, branch coverage, and multiple condition coverage.  
	It is considered one of the most rigorous coverage criteria. 
	The empirical results of previous studies~\cite{yu2006comparison,woodward2006relationship,gay2016effect,ahishakiye2021mc} indicate that it is more effective in terms of fault detection as compared to other coverage criteria.

	Model-based testing (MBT) facilitates automating various activities of software testing systematically. 
	It is a common practice of the avionics industry for the automated testing of safety-critical applications~\cite{sartajtesting}. 
    In MBT, creating models for the system under test (SUT) often involves using system specifications. 
    One of the methods is to develop these models using the Unified Modeling Language (UML)~\cite{uml} and specify constraints on these models using the Object Constraint Language (OCL)~\cite{ocl}.
	Both UML and OCL are standard languages defined by Object Management Group (OMG)\footnote{http://www.omg.org/} and are widely adopted by numerous software industries. 
	The models that are developed in this way along with the constraints are used for automating various testing activities. 
	For example, OCL constraints on UML models can be used for generating test data~\cite{ali2013generating, soltana2020practical}.

	To achieve MC/DC coverage of OCL constraints during test data generation, a large amount of effort and cost is involved. 
	Manual generation of test data for MC/DC is a tedious task. 
	The verification and validation objectives defined by international safety standards cannot be achieved by manual generation of test data for the MC/DC criterion. 
	Therefore, automation of different testing activities is mandatory.

    The research work presented in this paper was conducted in collaboration with a large avionics systems industry as a part of the project for the National Center of Robotics and Automation (NCRA). Their main goal is to comply with international safety standards, i.e., DO-178C while testing avionics systems of manned and unmanned aircraft.
    International safety standards like DO-178C recommend coverage of requirements at both system and source code levels using stringent coverage criteria, e.g., MC/DC. 
    OCL constraints specified on models at the system level are ultimately converted into conditions at the source code level~\cite{hemmati2018evaluating}. 
    Therefore, generating test data from OCL constraints targeting MC/DC coverage enables compliance with DO-178C test obligations, which is an important consideration of the avionics industry. 
    The available techniques targeting OCL-based MC/DC test data generation highlight the need for optimizing resource and time~\cite{hemmati2018evaluating}. 
    Our goal, in this context, is to devise an efficient strategy for search-based MC/DC test data generation using OCL constraints specified on models.

	In this paper, we propose an effective way to automate MC/DC test data generation during model-based testing.
	For this purpose, we develop a strategy that utilizes case-based reasoning (CBR)~\cite{cbr} and range reduction heuristics designed to solve MC/DC-tailored OCL constraints. 
	We performed an empirical study to compare our proposed strategy for MC/DC test data generation using CBR, range reduction, both CBR and range reduction, with an original search algorithm, and random search. 
    Additionally, we conducted an experiment to compare our proposed CBR and range reduction methods with two constraint solvers, namely \umltocsp~\cite{cabot2014verification} and \pledge~\cite{soltana2020practical}.
	We performed a large-scale experiment using 129 OCL constraints from six case studies of varying nature and sizes. 
    Among the six, one represents a Ground Control Station that is used to remotely control Unmanned Aerial Vehicles. 
    The other two case studies, EU-Rental~\cite{frias2003eu} and the Royal and Loyal~\cite{warmer2003object,oclrepo} are widely recognized in OCL literature as academic benchmarks~\cite{khan2019aspectocl}. 
    The fourth case study represents the UML meta-model of State Machines~\cite{uml}. 
    The fifth case study is an artificial (Seat Belt System) that we developed inspired by various industrial case studies that are published in the literature~\cite{ali2013generating,ali2015improving}. 
    The sixth case study is industrial which represents a Student Registration System.

	We evaluated our proposed strategy using the search algorithm, Alternating Variable Method (AVM). 
	The proposed strategy with CBR is termed \avmc, and with range reduction, it is termed \avmr. 
	We compared our strategy with the original AVM search, which is termed \avmo. 
	We used Random Search (RS) as a baseline for comparison. 
	We also analyzed the effect of the combination of CBR and range reduction (\avmrc, i.e., AVM with range reduction and CBR approach) and compared it with \avmc~and \avmr. 
	The experimental results showed that \avmo, \avmc, \avmr, and \avmrc~outperform RS. 
	The \avmc~and \avmr~performed significantly better when compared with \avmo~in terms of the ability to solve OCL constraints.
	The results of the comparison with \avmrc~indicated that the combination of CBR and range reduction (\avmrc) outperforms \avmr, \avmc, and \avmo. 
	Based on the results, the combined strategy (\avmrc) is considered the best option for solving MC/DC constraints. 
    Moreover, the results of the comparison with \umltocsp~and \pledge~showed that \avmrc~outperformed by solving a higher percentage of MC/DC constraints. However, in terms of time efficiency, \umltocsp~and \pledge~demonstrated better performance. 

    This paper is an extension of our previous conference paper published in the Proceedings of 11th International Symposium on Search-Based Software Engineering~\cite{sartaj2019search}. The specific contributions of this paper in addition to the conference version are as follows.
    We introduced a range reduction method into the MC/DC test data generation strategy to efficiently solve MC/DC-tailored OCL constraints. We significantly extended empirical evaluation in several ways. 
    First, we incorporated two new case studies comprising 45 OCL constraints. 
    Second, we added two new research questions (RQs). 
    One of the RQs analyzes our strategy with a combination of CBR and range reduction methods for all case studies. 
    Another RQ compares our proposed CBR and range reduction methods with two existing constraint solvers. 
    Third, we extended the experiment reported in the conference paper (evaluation with CBR) for the two new case studies and two new RQs. 
    Fourth, we evaluated our strategy with the range reduction method using six case studies, 129 OCL constraints, and all RQs.
    Lastly, we empirically evaluated our proposed methods with two existing constraint-solving approaches, i.e., \umltocsp~and \pledge.

	The remaining paper is structured as follows. 
	Section~\ref{bg} provides the background of model-based testing, MC/DC, search-based software engineering, OCL constraints solving, and case-based reasoning. 
	Section~\ref{re} presents a running example. 
	Section~\ref{app} describes our proposed strategy to generate test data for MC/DC using CBR and range reduction. 
	Section~\ref{exp} presents the empirical evaluation of our proposed strategy. 
	Section~\ref{rw} discusses the works related to our paper. 
	Finally, Section~\ref{con} concludes the paper.

	\section{Background} \label{bg}
This section provides the background of model-based testing, modified condition/decision coverage (MC/DC), search-based software engineering, OCL constraint solving, and case-based reasoning.

\subsection{Model-based Testing}

Model-based testing (MBT) supports testing automation using an abstract representation of the system under test (SUT) in the form of models~\cite{utting2010practical}. 
The process of MBT consists of five steps. 
In the first step, the specifications of the SUT are modeled using a modeling language, such as Unified Modeling Language (UML)~\cite{uml}. 
To model various aspects of the system, UML provides several modeling artifacts for different purposes. 
The modeling artifacts provided by UML are broadly categorized as structural models (e.g., a class diagram and a profile diagram) and behavioral models (e.g., a state machine and an activity diagram). 
The models developed in UML are augmented using a constraint specification language, i.e., Object Constraint Language (OCL)~\cite{ocl}. 
OCL is a widely used textual language for specifying constraints on UML models~\cite{khan2019aspectocl}.
Different UML models support the automation of several testing activities. 
For example, the UML state machine can be used to generate test sequences~\cite{iqbal2019model,sartaj2020cdst}. 
The second step is to select the coverage criteria for the test case selection. 
In the third step, the test case specification is defined based on the test selection criteria. 
In the fourth step, the test cases are generated using the model created in the first step, and the test case specification that is defined in the third step. 
Finally, in the fifth step, the executable test scripts for the SUT are created to execute the test cases generated in the previous step.

Several MBT techniques using different models, languages, and formalisms have been widely applied to test industrial applications such as mobile applications~\cite{karlsson2021model}, web applications~\cite{garousi2021model}, and cyber-physical systems~\cite{zafar2021model}. 
Similarly, various techniques using UML class diagrams and OCL constraints for test generation are also applied to several industrial systems~\cite{dadeau2019temporal,iqbal2015applying}. 
One of the techniques, similar to our work, is search-based which utilizes search algorithms to generate test data from OCL constraints specified on UML class diagrams~\cite{ali2013generating} (discussed in Section~\ref{oclsolve}). 
Search-based test data generation using OCL constraints has also been used for industrial systems~\cite{soltana2020practical,hemmati2018evaluating}. 
Solving industrial-scale OCL constraints with high complexity takes a lot of search budget and time as reported by Hemmati et al.~\cite{hemmati2018evaluating}, which is impractical for industrial systems.

\subsection{Modified Condition/Decision Coverage (MC/DC)}
MC/DC is a type of structural coverage criterion used in software testing.
This criterion subsumes other structural coverage criteria such as statement coverage, decision coverage, condition coverage, and condition/decision coverage~\cite{chilenski1994applicability}. 
Therefore, it is considered a stronger coverage criterion~\cite{ghani2009automatic}. 
It is a widely used criterion in the industry mainly in the avionics domain for testing safety and mission-critical applications.
It is also one of the main requirements of various international safety standards such as DO-178C~\cite{do178}.
\textcolor{black}{Chilenski and Miller~\cite{chilenski1994applicability} defined MC/DC as: ``Modified condition/decision coverage is a structural coverage criterion requiring that each condition within a decision is shown by execution to independently and correctly affect the outcome of the decision.''.}
For example, for a decision `$X$ $and$ $Y$', the possible combinations of each condition are TT, TF, FT, and FF. Out of these combinations, only TT, TF, and FT affect the decision, i.e., by changing $X/Y$ from T to F, or inversely, the decision outcomes change. These three combinations satisfy the MC/DC criterion and can be selected for testing. As the number of conditions in a decision increases, the total number of combinations increases by 2$^n$. This results in many combinations to test, which is infeasible for large-scale applications, e.g., safety-critical systems. MC/DC was designed to enable rigorous testing of safety-critical systems and comply with international safety standards~\cite{chilenski1994applicability}.

\subsection{Search-based Software Engineering (SBSE)}
In SBSE, a problem from the software engineering domain is formulated as a search problem, and then different search algorithms are used to solve that problem. 
A search problem formulation process typically consists of three steps~\cite{harman2001search}. 
The first step is to create a representation of the solution. 
The second step is to define a fitness function to evaluate the candidate solution during the search process.
Finally, the third step is to use operators (e.g., crossover) to evolve the candidate solution. 
The use of operators depends on the type of search algorithm (i.e., global or local).
The global search algorithms (e.g., genetic algorithms) use the whole population (with more than one individual) to search. 
The local search algorithms (e.g., hill climbing) start the search process using one individual. 
For global search algorithms like genetic algorithms, crossover, and mutation operators are used to generate offspring from two individuals. 
For local search algorithms like hill climbing, a mutation operator is needed to evolve an individual. 

\textbf{Alternating Variable Method (AVM).} It is a local search algorithm first introduced by Korel~\cite{korel1990automated}. It has been widely used to solve various software testing problems and has shown better performance than global search algorithms~\cite{mcminn2016avmf,korel1992dynamic,arcuri2009full,harman2009theoretical}.
AVM operates akin to hill climbing algorithms. It starts by initializing problem variables randomly. During the search process, AVM picks the first problem variable (keeping others constant), systematically changes variable value (e.g., $\pm$1), and observes fitness. If the current change leads to fitness improvement, it continues changing the variable value until a solution is found. Next, it selects the second problem variable and repeats the same procedure to find a solution. AVM search process continues until a solution for all problem variables is found or the search budget exceeds.

\subsection{OCL Constraint Solving}\label{oclsolve}
The process of solving an OCL constraint involves generating values for the attributes that can satisfy the overall constraint. 
To generate values that can be used as test data, Ali et al.~\cite{ali2013generating}, proposed a search-based approach. 
The approach takes a UML class diagram and OCL constraints as inputs, uses a search algorithm to solve the constraints, and generates test data in the form of an object diagram as output. 
For this purpose, a set of heuristics was proposed to guide the search algorithms. 
A heuristic to calculate branch distance tells ‘how far’ input values are from satisfying the constraint. 
Based on the fitness value, a search algorithm iteratively improves input values to satisfy the given OCL constraint. For example, consider that we want to solve a constraint \textit{C}: {\fontfamily{qcr}\selectfont context ClassX inv: self.x > \textcolor{black}{1}5} using the AVM algorithm. Suppose an initial random solution for {\fontfamily{qcr}\selectfont self.x > 15} is {\fontfamily{qcr}\selectfont $x=0$}. At the start, AVM evaluates this solution to analyze fitness. Since this solution is far from satisfying {\fontfamily{qcr}\selectfont self.x > 15}, AVM will increment ($x=1$) and decrement ($x=-1$) the solution value. As the incremented value ($x=1$) is close to satisfying {\fontfamily{qcr}\selectfont self.x > 15}, AVM will gradually improve the value of `x' toward better fitness (i.e., close to the solution) and continue until a solution is found or some stopping criteria are met. 
For example constraint \textit{C}, suppose after some iterations, the value of `x' is $x=31$ which satisfies {\fontfamily{qcr}\selectfont self.x > 15}. Now the solution is found and the AVM search process will stop. Finally, the solution for constraint \textit{C} is saved as an object diagram corresponding to the input class diagram.

\subsection{Case-based Reasoning}
Case-based Reasoning (CBR) introduces the concept of reusing previously solved solutions to solve the target problem with some identical characteristics. 
The process of CBR consists of four steps: (i) retrieve the previous solutions, (ii) select the solution to reuse, (iii) revise the old solution, (iv) and store the new solution. 
According to the CBR approach, a repository is maintained to store and retrieve solutions. 
Initially, the repository is empty. 
When the first problem is solved, its solution is stored in the repository along with its complete information which is considered a case. 
When the next problem is being solved, the repository is searched for solutions closest to the problem that can be reused. 
All possible closest solutions that have common parts are retrieved from the repository. 
The retrieved solutions are then evaluated individually to assess the closeness of the solution to the target problem. 
The closest solution is selected and reused to solve the target problem. 
When the target problem is solved using the previous solution, the newly generated solution is analyzed to determine whether it should be revised or not.
Finally, the new solution is stored in the repository. 
To illustrate the CBR process with an example, consider the OCL constraint \textit{C} and its solution $x=31$ (Section~\ref{oclsolve}), which is stored in a repository. Now consider the MC/DC constraint to solve is $C_m$: {\fontfamily{qcr}\selectfont context ClassX inv: self.x <= 15}. Since the repository has only one solution, this solution will be retrieved from the repository. AVM will use this solution (i.e., $x=31$), apply search to systematically increment/decrement the value of $x$, and try to satisfy {\fontfamily{qcr}\selectfont self.x <= 15}. After some iterations, suppose AVM finds a solution with value $x=8$. This solution will be saved in the repository.
	
\section{Running Example} \label{re}

\begin{figure}[h]
	\centerline{\includegraphics[width=12.5cm,height=8cm]{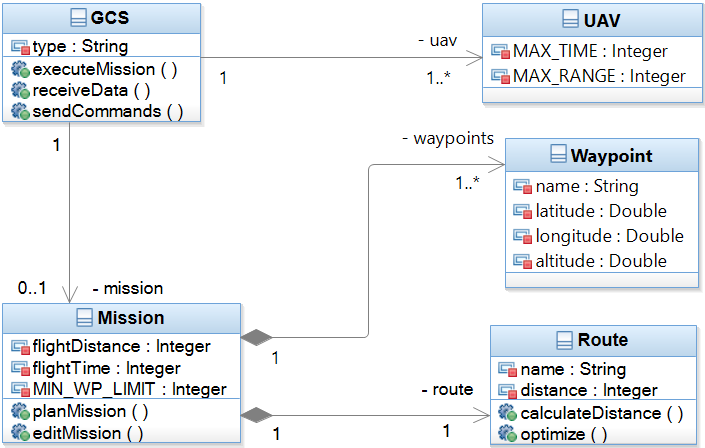}}
	\caption{An excerpt of GCS class diagram}
	\label{fig:gcs-ex}
\end{figure}

This section presents a running example containing a set of OCL constraints. 
For this purpose, we use an industrial case study representing a Ground Control Station (GCS).
A GCS is used to remotely operate an Unmanned Aerial Vehicle (UAV). 
One of the main functionalities of a GCS is to provide mission planning for the UAV.
Figure~\ref{fig:gcs-ex} shows an excerpt of the GCS class diagram with five classes. 
The main class is GCS which is associated with UAV and Mission classes. 
Each mission has one Route and multiple Waypoints. 
Listing~\ref{lst:re} shows 10 OCL constraints on various classes modeled for GCS. 
Among these constraints, five constraints (i.e., \textit{C1}, \textit{C2}, \textit{C3}, \textit{C7}, and \textit{C8}) are related to the UAV mission, and the remaining five constraints (i.e., \textit{C4}, \textit{C5}, \textit{C6}, \textit{C9}, and \textit{C10}) are modeled for route optimization. 

The constraint \textit{C1} demonstrates that for each mission defined using GCS, the UAV's flight time or flight distance must not exceed the maximum limit. \textit{C2} specifies that the number of waypoints in a mission must exceed the minimum threshold for the number of waypoints. \textit{C3} defines a limit on long missions (e.g., for monitoring or surveying) that mission must have 100 waypoints in addition to the minimum limit for waypoints. \textit{C4} defines pre-condition on route optimization operation in which route distance must be greater than the minimum distance. Similar to \textit{C4}, \textit{C5} and \textit{C6} define pre-conditions with margins of 1000 and 1500 feet respectively. \textit{C8} specifies UAV flight distance range with a margin of 100 meters. \textit{C9} defines a pre-condition on route optimization operation in which route distance must be greater than maximum range and input distance values should be valid. Lastly, \textit{C10} specifies a pre-condition on route optimization operation in which route distance with a margin must be greater than the total distance, and input distance values must be valid. \\

\begin{lstlisting}[label=lst:re, language=ocl, numbers=none,  framexleftmargin=2mm, xleftmargin=4mm, caption=OCL constraints from the GCS case study, linewidth=15.7cm]
--For a mission, either flight time should be less than the UAV's maximum endurance or distance should be less than the UAV's maximum range
C1: context GCS inv: self.mission.oclIsUndefined()=false and 
		(self.mission.flightTime<self.uav.MAX_TIME or 
			self.mission.flightDistance<self.uav.MAX_RANGE)
--The number of waypoints in a mission should be greater than the minimum number of waypoints
C2: context GCS inv: self.mission.waypoints>self.mission.MIN_WP_LIMIT
--For a survey mission, the number of waypoints should be greater than the minimum limit, and an additional 100 waypoints
C3: context GCS inv: self.mission.waypoints>self.mission.MIN_WP_LIMIT+100
--Route distance should be greater than the minimum distance
C4: context Route::optimize(in minDist : Integer, in maxDist : Integer)
      pre: self.distance>minDist
--Route distance with a 1000 ft. margin should be greater than the minimum distance
C5: context Route::optimize(in minDist : Integer, in maxDist : Integer)
      pre: self.distance+1000>minDist
--Route distance with a 1500 ft. margin should be greater than the total minimum and maximum distance
C6: context Route::optimize(in minDist : Integer, in maxDist : Integer)
      pre: self.distance+1500>minDist+maxDist
--Flight distance should be greater than 100 and less than 5000 meters
C7: context GCS inv: self.mission.flightDistance>100 
			and self.mission.flightDistance<5000
--Flight distance should be greater than MIN_RANGE+100 and less than MAX_RANGE-100
C8: context GCS inv: self.mission.flightDistance>self.uav.MIN_RANGE+100 and self.mission.flightDistance>self.uav.MAX_RANGE-100
--Route distance and maximum distance should be greater than the maximum range and minimum distance respectively
C9: context Route::optimize(in minDist : Integer, in maxDist : Integer)
      pre: self.distance>self.MAX_RANGE and maxDist>minDist
--Route distance with a 1500 ft. margin should be greater than the total minimum and maximum distance. Also, the minimum distance should be less than the maximum distance.
C10: context Route::optimize(in minDist : Integer, in maxDist : Integer)
      pre: self.distance+1500>minDist+maxDist and minDist<maxDist
\end{lstlisting}

	\section{Strategy to Generate Test Data for MC/DC} \label{app}
Generating test data for MC/DC requires the generation of multiple solutions for an OCL constraint corresponding to the MC/DC combinations. 
For example, consider a constraint with a predicate \textit{p} $ \lor $ \textit{q}, where \textit{p} and \textit{q} are the clauses. 
To satisfy this predicate, there are four possible combinations for both \textit{p} and \textit{q}, i.e., \textit{TT}, \textit{TF}, \textit{FT}, and \textit{FF}. 
For this example, the combinations according to the MC/DC criterion are \textit{TF}, \textit{FT}, and \textit{FF}.

\begin{figure*}[!t]
    \centerline{\includegraphics[width=12.9cm,height=11.4cm]{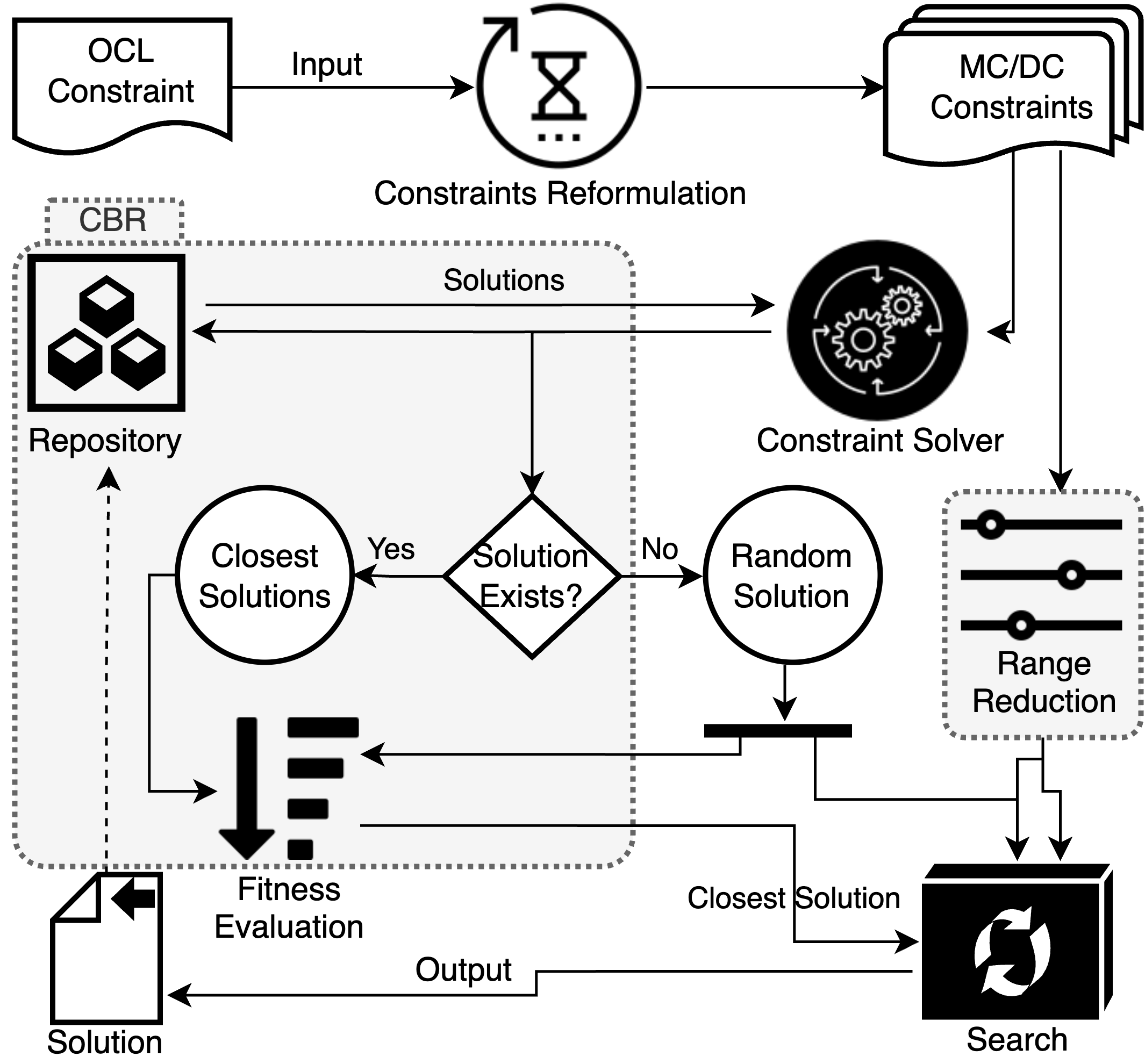}}
	\caption{An overall view of the proposed MC/DC strategy}
	\label{fig:mcdc-app}
\end{figure*}

Figure~\ref{fig:mcdc-app} shows an overall view of the proposed strategy to generate test data for MC/DC. 
As an initial step, the strategy takes an OCL constraint as input and reformulates it to obtain a list of MC/DC-tailored OCL constraints. 
The process of constraint reformulation involves handling the negation of logical expressions and the negation of OCL collections. 
The next step is to solve MC/DC-tailored OCL constraints using a search-based constraint solver. 
For this purpose, we propose two methods (as highlighted in Figure~\ref{fig:mcdc-app}).  
The first method applies the concept of case-based reasoning (CBR) to optimize the process of constraint solving for the MC/DC-based constraints. 
The idea of CBR is to maintain a repository for solutions to reuse the previous solution while solving the current problem instead of solving it from scratch. 
Whenever a new MC/DC constraint is being solved, the repository is examined for the closest previous solutions. 
If no previous solution exists, the constraint solver generates a random solution and starts the search. 
If one or more closest previous solutions exist, the fitness of each previous and random solution is checked. 
This results in the closest solution that can be used as an initial seed of a search algorithm. 
The second method reduces the range of primitive attributes in the MC/DC constraints and generates random solutions within reduced ranges. 
These solutions are used by a search algorithm as a starting point to systematically change values during the search process. 
The aim is to minimize overall search steps which leads to further improvement in solving MC/DC constraints.
Both methods can be used individually and collectively. 
At the end of the search process, the solution generated for the MC/DC constraint is stored in the repository for future reuse. 
In the following, we elaborate on the steps of our strategy.

\subsection{OCL Constraints Reformulation for MC/DC}
To apply the MC/DC criterion, a number of truth value combinations are required corresponding to the predicate of an OCL constraint. 
For this purpose, we use the pair-table approach presented by Chilenski and Miller~\cite{chilenski1994applicability}. 
The pair-table approach provides a systematic way of identifying the combinations required for MC/DC.
Consider an example OCL constraint \textit{C1} specified on \textit{GCS} class (Listing~\ref{lst:re}).
We transform the predicate of \textit{C1} to an equivalent Boolean expression and assign a unique identifier to each clause of the predicate. 
Resultantly, the Boolean expression obtained for this example is $ p \land (q \lor r) $.

The next step is to create a pair table.
For the example constraint \textit{C1}, Table~\ref{tab:pt} shows the pair table for the transformed Boolean expression. 
Column 2 of the table shows all possible truth values corresponding to the predicate and column 3 contains the outcome of the predicate. 
The pair table contains columns such as \textit{p}, \textit{q}, and \textit{r} corresponding to each clause. 
The values for the column represent the potential pairs that fulfill the criteria of MC/DC combination selection. 
For example, row 2 corresponding to the column \textit{p} in the table has a value of 6, which suggests that the combinations in row 2 and row 6 are potential pairs. 
Similarly, the combinations in rows 2 and 4 are potential pairs for clause \textit{q}.

\begin{table}[!htb]
		\centering
		\caption{Example pair table for \textit{C1}}
		\label{tab:pt}
		\begin{tabular}{|g|g|g|g|g|g|}
			\hline
			\rowcolor{white}
			\textbf{\#} & \textbf{pqr} & \textbf{Result} & \textbf{p} & \textbf{q} & \textbf{r} \\
			\hline
			\rowcolor{white}
			1 & TTT & T & 5 &  &  \\  
			\textbf{2} & \textbf{TTF} & \textbf{T} & \textbf{6} & \textbf{4} &  \\  
			\textbf{3} & \textbf{TFT} & \textbf{T} & \textbf{7} &  & \textbf{4} \\  
			\textbf{4} & \textbf{TFF} & \textbf{F} &  & \textbf{2} & \textbf{3} \\  
			\rowcolor{white}
			5 & FTT & F & 1 &  &  \\  
			\rowcolor{Gray}
			\textbf{6} & \textbf{FTF} & \textbf{F} & \textbf{2} &  &  \\  
			\rowcolor{white}
			7 & FFT & F & 3 &  &  \\  
			\rowcolor{white}
			8 & FFF & F &  &  &  \\  
			\hline
		\end{tabular}
\end{table}
\begin{table}[!htb]
		\centering
		\caption{Heuristics for the negation of OCL collection operations}
		\label{tab:neg}
		\begin{tabular}{|l|l|l|}
			\hline
			\textbf{\#} & \textbf{Collection Operation} & \textbf{Negation} \\
			\hline
			1 & \textit{forAll(p)} & \textit{exists(not p)} \\  
			2 & \textit{exists(p)} & \textit{forAll(not p)} \\  
			3 & \textit{one(p)} & \textit{select(p)$\rightarrow$size()$<$$>$1} \\  
			4 & \textit{includes(N)} & \textit{excludes(N)} \\  
			5 & \textit{select(p)} & \textit{select(not p) or reject(p)} \\  
			6 & \textit{select(p)$\rightarrow$isEmpty()} & \textit{select(p)$\rightarrow$notEmpty()} \\  
			7 & \textit{reject(p)$\rightarrow$isEmpty()} & \textit{reject(p)$\rightarrow$notEmpty()} \\  
			8 & \textit{select(p)$\rightarrow$size()=C} & \textit{select(p)$\rightarrow$size()$<$$>$C} \\    
			9 & \textit{select(p)$\rightarrow$size()$<$C} & \textit{select(p)$\rightarrow$size()$>$=C} \\  
			10 & \textit{select(p)$\rightarrow$size()$>$C} & \textit{select(p)$\rightarrow$size()$<$=C} \\  
			\hline
		\end{tabular}
\end{table}

To identify MC/DC combinations, \textcolor{black}{we use the pair table constructed in the previous step. We select minimum subsets of pairs that cover both truth combinations (T/F) of all clauses. }
For the given example, the minimum subsets of pairs are $\{$2,6$\}$, $\{$2,4$\}$, and $\{$3,4$\}$ (Table~\ref{tab:pt}). 
The MC/DC combinations corresponding to these pairs are TTF, TFT, TFF, and FTF. 
Using these MC/DC combinations, the predicate of the OCL constraint is reformulated. 
In this process, a clause is negated for each false value in the MC/DC combination. 
For example, for the combination TTF, the predicate $ p \land (q \lor r) $ needs to be reformulated with a negated \textit{r}, i.e. $ p \land (q \lor \neg r) $.

After obtaining MC/DC combinations, one way is to tune a constraint solver to 
internally handle clause negations and logical operations. However, tuning a constraint solver is insufficient for clauses with OCL collection operations and logical operations with `xor' and `implies', as described in the following subsections with examples. Considering this, we reformulate an OCL constraint by handling OCL negations and logical operations for each MC/DC combination. The reformulated MC/DC does not require tuning a particular constraint solver and can be used with any other constraint solver.

\subsubsection{Handling Negation for OCL}
OCL constraints are typically composed of clauses with relational operators and clauses with OCL collection operations. 
The clause with relational operators can be negated either by adding \textit{not} operation before the clause or by inverting the relational operator, e.g., the negation of relational operator $<$ is $>=$. 
The clause with OCL collection operations cannot be negated straightforwardly because their negation directly affects the elements of the collection. 
Table~\ref{tab:neg} shows heuristics for the negation of OCL collection operations. 
For the negation of \textit{forAll(p)}, we need to negate the predicate (i.e., \textit{p}) for the entire collection and change the collection operation to \textit{exists()}. 
The final negation form of \textit{forAll(p)} is \textit{exists(not p)} (Table~\ref{tab:neg}, row \# 1). 
Similar is the case with the negation of \textit{exists(p)}, i.e., we need to negate the predicate \textit{p} and change the collection operation to \textit{forAll()}. 
The final negation form of \textit{exists(p)} is \textit{forAll(not p)} as shown in (Table~\ref{tab:neg}, row \# 2). 
The negation of collection operation \textit{one(p)} is different. 
The operation \textit{one(p)} requires exactly one instance for which the predicate evaluates to \textit{true}. 
Thus, the negation \textit{one(p)} operation is \textit{select(p)} with \textit{size()} operation in which the size is not equal to one (Table~\ref{tab:neg}, row \# 3). 
The negation of \textit{includes(N)} operation changes it to \textit{excludes(N)} and vice versa (Table~\ref{tab:neg}, row \# 4). 
In the case of \textit{select(p)} operation, it can either be changed to \textit{select(not p)} or \textit{reject(p)} (Table~\ref{tab:neg}, row \# 5). 
Similar is the case with \textit{reject(p)} operation. 
When the collection operations \textit{select(p)} and \textit{reject(p)} are used with \textit{isEmpty()} operation, the negation is further applied to \textit{isEmpty()} by converting it to \textit{notEmpty()} (Table~\ref{tab:neg}, rows: 6 and 7). 
Similarly, when \textit{select(p)} operation is used with \textit{forAll(p)}, \textit{exists(p)}, or \textit{one(p)} operation, the negations of these operations are also applied. 
Furthermore, when \textit{select(p)} is used with \textit{size()} operation, this means the predicate evaluates to \textit{true} for only a specified number of instances of the collection. 
Therefore, in the case of \textit{select(p)} with \textit{size()} operation, the relational operator used with \textit{size()} is negated (Table~\ref{tab:neg}, rows: 8-10).

\begin{lstlisting}[label=lst:2, language=ocl, caption=Reformulated MC/DC constraints for \textit{C1}, linewidth=15.7cm]
C1: context GCS inv: self.mission.oclIsUndefined()=false and 
			(self.mission.flightTime<self.uav.MAX_TIME and 
	   		self.mission.flightDistance>=self.uav.MAX_RANGE)         --TTF
C2: context GCS inv: self.mission.oclIsUndefined()=false and 
			(self.mission.flightTime>=self.uav.MAX_TIME and 
	   		self.mission.flightDistance<self.uav.MAX_RANGE)          --TFT
C3: context GCS inv: self.mission.oclIsUndefined()=false and 
			(self.mission.flightTime>=self.uav.MAX_TIME and 
	   		self.mission.flightDistance>=self.uav.MAX_RANGE)         --TFF
C4: context GCS inv: self.mission.oclIsUndefined()=true and 
			(self.mission.flightTime<self.uav.MAX_TIME and 
	   		self.mission.flightDistance>=self.uav.MAX_RANGE)         --FTF
\end{lstlisting}

\subsubsection{Logical Operations Reformulation}
To obtain the exact truth values that represent MC/DC combinations derived from the pair table, only negating various clauses is insufficient. 
The reason is that the heuristics (presented by Ali et al.~\cite{ali2015improving}) for logical operations such as \textit{or}, \textit{xor}, and \textit{implies} are developed to improve the search by solving the `easiest' clause first. 
For example, in the case of predicate ``\textit{A} or \textit{B}'', the heuristic to calculate the branch distance~\cite{ali2015improving} is: If (\textit{d(A) }$<$=  \textit{d(B)}) then solve \textit{A} otherwise solve \textit{B} clause. 
If the MC/DC combination requires the negation of clause \textit{B}, it is possible that the negation of \textit{B} is hard to solve as compared to clause \textit{A}. 
In this case, the existing heuristics lead towards solving clause \textit{A}. 
Therefore, it is not certain that the generated test data reflects the negation of \textit{B}. 
The same is the case with \textit{xor} and \textit{implies} heuristics. 
To handle this, our strategy performs a second pass to reformulate the logical operations so that the generated test data directly reflects the MC/DC combination. 

Our strategy converts the logical operation `or' to `and' between only the negated clauses.
In the case of `implies' logical operation, our strategy first converts this into a logically equivalent expression with basic logical operations such as \textit{and}, \textit{or}, and \textit{not}. 
For example, the logically equivalent expression of the predicate \textit{a} implies \textit{b} is: \textit{not} (\textit{a}) or \textit{b}. 
When the logically equivalent expression is obtained, each `or' logical operation between the negated clauses is changed to `and'.
Similar is the case with the `xor' operation. 
First, the expression with the `xor' logical operation is converted to its logically equivalent expression.
For example, the logically equivalent expression of the predicate \textit{a} xor \textit{b} is:\textit{ }(\textit{a }or \textit{b}) and \textit{not} (\textit{a }and \textit{b}). 
The operation `or' is changed to `and' in the obtained logically equivalent expression. 

For the example constraint \textit{C1} (Listing~\ref{lst:re}), four MC/DC combinations (i.e., TTF, TFT, TFF, and FTF) are obtained from the pair table. 
After reformulation of OCL constraints according to MC/DC combinations, the obtained MC/DC constraints are shown in Listing~\ref{lst:2}. 
The MC/DC constraint \textit{C1} in Listing~\ref{lst:2} corresponds to the combination TTF. 
It has the third clause negated and the logical operation converted from \textit{or} to \textit{and} operation. 
Similarly, the remaining MC/DC constraints \textit{C2}, \textit{C3}, and \textit{C4} are obtained after reformulating according to MC/DC combinations.

\subsubsection{Identifying Conflicting MC/DC Constraints}
The MC/DC constraints obtained after reformulation are not always solvable due to infeasible combinations~\cite{rajan2008effect}. 
For example, if the combination is FF for the predicate $ x\geq10$ or  $x\leq 25 $, it is an infeasible combination because the resulting constraint will contain conflicting clauses. 
A constraint is considered as conflicting if two or more clauses with primitive attributes are dependent. 
Two or more clauses can be dependent on one another if the same primitive attribute is used in various clauses. 
For example, both clauses of the predicate $ x\geq10$ or  $x\leq 25 $ are dependent because they have a common attribute \lq x\rq.

Since a conflicting MC/DC constraint cannot be solved, our strategy identifies the conflict by finding the dependent clauses first. 
The dependency among clauses is identified by examining the use of common primitive attributes. 
If two or more clauses are dependent, we analyze the fitness values for half of the total search budget. 
For a non-conflicting constraint, its fitness values should improve with the increase in fitness evaluations regardless of constraint complexity. 
In the case where fitness values do not improve for a specified search budget, the constraint is considered conflicting and the search process is terminated. 
For the confirmation, we need to manually analyze whether the constraint was conflicting or not. 
Besides conflict due to dependent clauses, a constraint with independent clauses can be conflicting. In our approach, such conflict can be identified by analyzing the fitness values. However, tools like QMaxUSE~\cite{wu2023qmaxuse}, or other techniques~\cite{clariso2022managing,liffiton2008algorithms} can be used to detect such conflicts. 

After the reformulation of all OCL constraints for the MC/DC criterion, the next step is to solve MC/DC constraints. 
A straightforward approach is to solve each MC/DC constraint individually (e.g., using the approach by Ali et al.~\cite{ali2013generating}). 
The problem with this approach is that it is a time-consuming task. 
It can be observed from the reformulation step that when a single OCL constraint is reformulated to apply MC/DC, one constraint is converted to multiple MC/DC constraints (as shown in Listing~\ref{lst:2}).
According to our experience with industrial constraints~\cite{iqbal2015applying}, it is very time-consuming to solve industrial-scale OCL constraints with different complexity levels~\cite{ali2014insights}. 
It can take up to 15 minutes to solve a single OCL constraint.
Thus, solving each MC/DC constraint separately can consume a lot of computing resources (such as time and search budget~\cite{hemmati2018evaluating}), especially, when the number of clauses in an OCL constraint increases. 
To handle this problem, we propose two methods to improve the search for solving MC/DC constraints. 
In the following sections, we discuss both methods. 

\subsection{Solving MC/DC Constraints using Case-based Reasoning}

Analyzing the example MC/DC constraints as shown in Listing~\ref{lst:2}, we notice that constraints \textit{C1}, \textit{C2}, and \textit{C3} have the same first clause, and constraints \textit{C3} and \textit{C4} have an identical third clause. 
Whenever MC/DC is applied, the resulting MC/DC constraints exhibit this behavior. 
So, rather than solving each MC/DC constraint individually, we can reuse a component of already solved constraints. 
Therefore, when the previous solution is reused, it contains the solution for an already solved clause(s). 
As a result, the search algorithm starts the search from a partially solved individual rather than a random individual. 
In this way, finding a solution for a constraint becomes much easier compared to solving a constraint from scratch.

For this purpose, we adopt the concept of case-based reasoning (CBR)~\cite{cbr} to reuse the previous solution while solving MC/DC constraints. 
The CBR approach provides a well-suited method to reuse solutions of previously solved problems while solving a current problem with a part similar to the previously solved problems.

To use the CBR approach for solving MC/DC constraints, first, we define some terminologies. 
Let $C=\{C_1, C_2, C_3,\dots, C_n\}$ be the set of MC/DC constraints to solve and the previously solved solutions in the repository are $S=\{S_1, S_2, S_3,\dots, S_{n-1}\}$, where \textit{n} is the total number of MC/DC constraints. 
When the MC/DC constraint $C_1$ is solved, its solution $S_1$ is stored in a repository so that it can be reused. 
Initially, the repository is empty and successively filled with each constraint's solution till solving the last constraint. 
After the last constraints solution, the repository is utilized for a set of MC/DC constraints. 
Therefore, the total number of solutions ($S_{n-1}$) in the repository is one less than the total number of MC/DC constraints. 
We use the MC/DC constraint, the predicate, and the resultant data to store a solution in the repository. 
Thus, a solution is represented as a tuple: $S_k=(C_k, P_k, data)$. 
Here $S_k$ denotes a solution representation in the repository consisting of an MC/DC constraint ($C_k$), the predicate ($P_k$), and the resultant \textit{data}. A predicate is a part of an MC/DC constraint. Keeping both the MC/DC constraint and its predicate in a solution representation is required when multiple constraints on one class context are combined. Therefore, the purpose of such a solution representation is to uniquely identify solution data corresponding to a constraint and to use its predicate for distance calculation. 
Moreover, $P_s=\{P_{s_1}, P_{s_2}, P_{s_3},\dots,\ P_{s_{n-1}}\}$ are the list of predicates of MC/DC constraints for which the repository contains solutions. 
Each predicate in the repository consists of several clauses and it is represented as $P_{s_i}=\{c_{s_1}, c_{s_2}, c_{s_3},\dots, c_{s_m}\}$. 
For the MC/DC constraint being solved, the target predicate is represented as $P_t=\{c_{t_1}, c_{t_2}, c_{t_3},\dots, c_{t_m}\}$, where \textit{m} is the total number of clauses in a predicate. 

Our strategy measures the similarity between the target predicate and the already solved predicate to identify a previous solution that is appropriate to reuse.  
For this purpose, we calculate the similarity score between the two predicates using Equation~\ref{eq1}. 
Since a predicate consists of multiple clauses, we calculate the similarity between individual clauses of the target predicate and the predicates representing previous solutions. 
In this equation, $d\left(c_{t_j}, c_{s_ij}\right)$ is the similarity score between a target clause ($c_{t_j}$) and the clauses of previously solved constraints ($c_{s_ij}$) from the repository. 
If the similarity score between the two predicates is high, that previous solution is considered more appropriate to reuse.  
The similarity score between the two individual clauses is calculated by using Equation~\ref{eq2}.

\begin{equation}\label{eq1}
d\left(P_t,\ P_{s_i}\right)=\ \sum^m_{j=1}{d\left(c_{t_j},\ c_{s_ij}\right)}
\end{equation}

\begin{equation}\label{eq2}
  d\left(c_{t_j},\ c_{s_ij}\right) = \left\{ 
  \begin{array}{ l l }
    1 & \quad c_{t_j}=\ c_{s_ij} \\
    0 & \quad c_{t_j}\neq\ c_{s_ij}  
  \end{array}
\right.
\end{equation}

According to Equation~\ref{eq2}, if the two clauses are identical, the similarity score between them is 1 and 0 otherwise. 
For example, the similarity score between MC/DC constraints \textit{C3} and \textit{C4} (shown in Listing~\ref{lst:2}) is one, which is calculated as follows.
\begin{center}
$d\left(P_4,\ P_{3_1}\right)=\ d\left(c_{4_1},\ c_{3_11}\right) + d\left(c_{4_2},\ c_{3_12}\right) + d\left(c_{4_3},\ c_{3_13}\right)$\\
$d\left(P_4,\ P_{3_1}\right)= 0 + 0 + 1=1$
\end{center}
Using Equations~\ref{eq1} and~\ref{eq2}, the similarity score of each solution in the repository with the target predicate is calculated. 
Based on the similarity score, we get the possible set of closest solutions.

\textbf{Reusing the Previous Solution.}
Typically, a search algorithm starts the search process using a random initial seed. 
To optimize the search for reusing the closest previous solutions, the initial seed needs to be changed. 
Once a set of possible closest previous solutions is obtained, we need to select one nearest solution as an initial seed to start the search. 

Let $N_s=\{N_{s_1}, N_{s_2}, N_{s_3}, \dots , N_{s_x}\}$ be the list of \textit{x} number of nearest possible solutions that are obtained after calculating the similarity score. 
In the case where the same similarity score of two or more nearest solutions in $N_s$ is the same, we calculate the fitness of each nearest solution and then select the solution with minimum fitness. 
To calculate fitness, we use the fitness functions proposed by Ali et al.~\cite{ali2013generating} for solving OCL constraints. 
The fitness function determines how far the solution of the constraint is from evaluating to \textit{true}. 
For example, consider a constraint is C: x$>$0 that has two possible candidate solutions, i.e.,  $S_1$: x=-125 and $S_2$: x=0. 
The solution $S_1$ is far from satisfying \textit{C} as compared to the solution $S_2$. 
Therefore, the fitness value of $S_2$ will be less as compared to $S_1$ because it is very close to satisfying the constraint.

The fitness value of each nearest solution in $N_s$ corresponding to the target MC/DC constraint is $f_{s_n}=\{f_{s_1}, f_{s_2}, f_{s_3},\dots, f_{s_x}\}$. 
Let $S_c$ be the closest solution from the set $N_s$ with minimum fitness $f_{s_c}$ which is calculated as $f_{s_c}=\mathrm{min}(f_{s_n})$. 
Now, according to the actual behavior of a search algorithm, a random initial seed is generated. 
Let $S_r$ be a random individual (initial seed) with fitness value $\ f_{s_r}$. 
The nearest previous solution is determined by using Equation~\ref{eq3}. 

\begin{equation}\label{eq3}
  S_T\mathrm{=}  \left\{ 
  \begin{array}{ l l }
    S_c & \quad f_{s_c}\mathrm{<}f_{s_r} \\
    S_r & \quad f_{s_c}\geq f_{s_r}  
  \end{array}
\right.
\end{equation}

According to Equation~\ref{eq3}, if the fitness value $f_{s_c}$ of the closest previous solution $S_c$ from $N_s$ is less than that of the random solution $S_r$, the solution $S_c$ is selected. 
Otherwise, the search is started from a random solution $S_r$. 
Since it is possible that a randomly generated initial seed can be an appropriate option to start the search, we also check the fitness value of the random individual. 

\subsection{Solving MC/DC Constraints using Range Reduction}

To solve an OCL constraint, an important decision is to select an initial range for the Integer and Real type attributes.
The range of values for Integer and Real types is very large.
For example, the Integer number has values ranging from $-2^{31}$ to $+2^{31}$.
Due to the large range, a subset of the entire range (e.g., -10000 to 10000) is selected.
This range remains the same for all Integer/Real type attributes used in various OCL constraints. 
The main problem with selecting a fixed range for all attributes is that the range may not be suitable for some attributes. 
Some attributes either require a very small range of values or a large range of values.  
For example, the \textit{age} attribute of a person requires a range of values from 0 to 100. Similarly, the \textit{price} attribute of a car requires a range of values from 10000 to 500000.
Therefore, the use of a fixed range for all attributes to solve OCL constraints increases the search time. 
The constraint may not be solved within the specified search budget which leads to degrading the performance. 

We address the problem of using a fixed range for all attributes.
We present heuristics to select the range for a specific constraint with Integer and Real type of attributes.
In the case of Boolean and Enumeration type of attributes, the range is already small. 
The Boolean attribute can have only two values in its search domain, i.e., \textit{true} and \textit{false}. 
The Enumeration type of attribute contains a fixed number of Enumeration literals. 
For example, the enumeration for weekdays contains only seven Enumeration literals.
In addition, constraint clauses may contain OCL collection operations and user-defined operations. However, such operations are rarely used in industrial constraints~\cite{iqbal2015applying,ali2014insights}. Moreover, the search space in such operations is typically small and can be solved easily without range reduction.
Therefore, we focus on two commonly used primitive types, i.e., Integer and Real. 
In the following, we discuss heuristics to select the range based on the nature of OCL constraints. 

An OCL constraint consists of several clauses.
A clause contains an OCL expression.
In the case of Integer and Real type attributes, a clause is represented by an expression with an attribute name, a relational operator, and a constant.
For example, OCL constraint C2 as shown in Listing~\ref{re} contains an invariant with a single clause, i.e., \lq self.mission.waypoints $>$ self.mission.MIN\_WP\_LIMIT\rq.
In this clause, the \lq waypoints\rq\space is the attribute name, the relational operator is \lq$>$\rq, and the constant is MIN\_WP\_LIMIT. 
In the following, we present heuristics to reduce the range for different types of clauses with Integer and Real type attributes. 
\textcolor{black}{First, we introduce cases involving single clause constraints (Section~\ref{sc}), followed by a discussion of the cases involving multiple clauses joined with Boolean operators (Section~\ref{mc}). 
At the single clause level, cases in Section~\ref{sc} can be considered special cases of those presented in Section~\ref{mc}. 
We describe each case individually to ensure a more in-depth understanding of every scenario. 
}

\subsubsection{Constraint with Single Clause}\label{sc}
In the case when a constraint consists of a single clause,  at least one attribute with one constant is present in the clause.
Let \textit{A} be the attribute and \textit{c} be the constant used in a clause.
The minimum and maximum values for the range of the attribute \textit{A} are calculated using Equations~\ref{eq4} and~\ref{eq5} respectively.
These equations present range calculation formulas corresponding to positive, negative, and zero values of \textit{c}.

\begin{equation}\label{eq4}
  A_{min}\mathrm{=}  \left\{ 
  \begin{array}{ l l }
    \textcolor{black}{0} & \quad c\mathrm{>}0 \\
    \textcolor{black}{2 * c} * s_f & \quad c\mathrm{<}0 \\
    -1 * s_f & \quad c\mathrm{=}0 
  \end{array}
\right.
\end{equation}

\begin{equation}\label{eq5}
  A_{max}\mathrm{=}  \left\{ 
  \begin{array}{ l l }
    \textcolor{black}{2 * c}  * s_f & \quad c\mathrm{>}0 \\
    \textcolor{black}{0} & \quad c\mathrm{<}0 \\
    1 * s_f & \quad c\mathrm{=}0 
  \end{array}
\right.
\end{equation}

In Equations~\ref{eq4} and~\ref{eq5}, $s_f$ is a scaling factor.
Its values are positive non-zero integers, i.e., $s_f = \{1, 2, 3, \dots\}$.
The values of $s_f$ determine the scale of minimum and maximum values for specifying the range. 
\textcolor{black}{
Initially, domain experts must select a scaling factor value considering the constraints' complexity and case study specifications, which is a one-time effort. 
This predetermined scaling factor value should be set at the beginning of the search process.
It can then be utilized for subsequent generation tasks.
It is important to note that a higher scaling factor will result in a correspondingly higher range.
}

For example, consider a constraint \textit{C2} as shown in Listing~\ref{lst:re}. 
The constraint \textit{C2} contains an invariant with a single clause, i.e., \lq self.mission.waypoints $>$ self.mission.MIN\_WP\_LIMIT\rq.
In this clause, \textit{waypoints} is the attribute (\textit{A}) and the constant \textit{c} is MIN\_WP\_LIMIT.
If the value of MIN\_WP\_LIMIT is 10, the constant \textit{c} is 10, i.e., c=10.
If the value of the scaling factor is set to 1, i.e., $s_f = 1$, the minimum and maximum values are calculated as $ A_{min} = \textcolor{black}{0} $ and $ A_{max} =  \textcolor{black}{2 * 10 * 1 = 20} $.
Therefore, the range of the attribute (\textit{A}) with $s_f = 1$ is $[0, 20]$.
If the value of the scaling factor is set to 2, i.e., $s_f = 2$, the minimum and maximum values are calculated as  $ A_{min} = \textcolor{black}{0} $ and $ A_{max} = \textcolor{black}{2 * 10 * 2 = 40} $. 
Therefore, the range of the attribute (\textit{A}) with $s_f = 2$ is $[0, 40]$.
As the value of the scaling factor increases, the range of the attribute increases.

\textbf{Case 1: Constant with arithmetic manipulation.}
In the case when a constraint consists of a single clause and the constants require arithmetic manipulation.
The heuristic for this case is to solve the constants expression according to the precedence of the arithmetic operator(s).
The resultant constant value obtained after solving the expression is used for range calculation using Equations~\ref{eq4} and~\ref{eq5}.

For example, consider a constraint \textit{C3} as shown in Listing~\ref{lst:re}.
The constraint \textit{C3} contains an invariant with a single clause, i.e., \lq self.mission.waypoints $>$ self.mission.MIN\_WP\_LIMIT+100\rq.
In this clause, \textit{waypoints} is the attribute (\textit{A}) and there are two constants i.e., $c_1 = $MIN\_WP\_LIMIT and $c_2 = 100$.
The constants expression contains an addition arithmetic operator (i.e., +).
If the value of constant MIN\_WP\_LIMIT is 20, the resultant constant after solving the arithmetic expression is $c = 120$. 
If the value of the scaling factor is set to 1, i.e., $s_f = 1$, the minimum and maximum values are calculated as $ A_{min} = \textcolor{black}{0} $ and $ A_{max} = \textcolor{black}{2 * 120 * 1 = 240} $.
Therefore, the range of the attribute (\textit{A}) with $s_f = 1$ is $[0, 240]$.
If the value of the scaling factor is set to 2, i.e., $s_f = 2$, the minimum and maximum values are calculated as $ A_{min} = \textcolor{black}{0} $ and $ A_{max} = \textcolor{black}{2 * 120 * 2 = 480} $. 
Therefore, the range of the attribute (\textit{A}) with $s_f = 2$ is $[0, 480]$.

\textbf{Case 2: Attribute/Variable on both sides of a relational operator.}
In the case when a clause only contains attributes/variables\footnote{The difference between an attribute and a variable is that an attribute is the property of class model whereas the variable is not present in the model but defined in OCL constraint.} on both sides of a relational operator and there is no constant involved in that clause.
To handle this case, the heuristic is to choose the smallest positive integer value randomly in the range [0, 100]. 
We define the range with values less than 100 so that the attributes/variables with a high difference will not lead to a large range, which will require more execution time of a constraint solver. 
The selected Integer value is considered as a constant \textit{c} to determine the range of both attributes or variables using Equations~\ref{eq4} and~\ref{eq5}.

For the example constraint \textit{C4} as shown in Listing~\ref{lst:re}, $ A_1 $ and $ A_2 $ represent the attribute \textit{distance} and the variable \textit{minDist} respectively. 
If the smallest positive integer value selected at random is 5, i.e., $c = 5$.
If the value of the scaling factor is set to 1, i.e., $s_f = 1$, the minimum and maximum values are calculated as  $ A_{1_{min}} = A_{2_{min}} = \textcolor{black}{0} $ and $ A_{1_{max}} = A_{2_{max}} = \textcolor{black}{2 * 5 * 1 = 10} $.
Therefore, the range of $ A_1 $ and $ A_2 $ with $s_f = 1$ is $[0, 10]$.
If the value of the scaling factor is set to 2, i.e., $s_f = 2$, the minimum and maximum values are calculated as  $ A_{1_{min}}  = A_{2_{min}} = \textcolor{black}{0} $ and $ A_{1_{max}}  = A_{2_{max}} = \textcolor{black}{2 * 5 * 2 = 20} $.
Therefore, the range of $ A_1 $ and $ A_2 $ with $s_f = 2$ is $[0, 20]$.

\textbf{Case 3: Attribute/variable on both sides of a relational operator and contain constant(s).}
In the case when a clause contains attributes/variables on both sides of a relational operator and there is constant(s) involved on either one side or both.
If there is one constant involved in a clause, the value of the constant \textit{c} is used to calculate the range using Equations~\ref{eq4} and~\ref{eq5}.
If there are more than one constants involved in a clause, the minimum value from constants is used as resultant constant \textit{c} to calculate the range using Equations~\ref{eq4} and~\ref{eq5}.

For example, consider a constraint \textit{C5} as shown in Listing~\ref{lst:re}.
The constraint contains a pre-condition with a single clause, i.e., \lq self.distance+1000 $>$ minDist\rq.
In this clause, \textit{distance} is the attribute (\textit{A}), the constant \textit{c} is 1000, and \textit{minDist} is the variable.
If the value of the scaling factor is set to 1, i.e., $s_f = 1$, the minimum and maximum values are calculated as $ A_{min} = A_{2_{max}} = \textcolor{black}{0} $ and $ A_{max} = A_{2_{max}} = \textcolor{black}{2 * 1000 * 1 = 2000} $.
Therefore, the range of the attribute (\textit{A}) with $s_f = 1$ is $[0, 2000]$.
If the value of the scaling factor is set to 2, i.e., $s_f = 2$, the minimum and maximum values are calculated as  $ A_{min} = A_{2_{max}} = \textcolor{black}{0} $ and $ A_{max} = A_{2_{max}} = \textcolor{black}{2 * 1000 * 2 = 4000} $.
Therefore, the range of the attribute (\textit{A}) with $s_f = 2$ is $[0, 4000]$.

\textbf{Case 4: Multiple attributes/variables on both sides of a relational operator.}
OCL constraints can contain clauses in which multiple attributes are used on both sides of a relational operator.
If there is no constant on either side of a relational operator, choose the smallest positive integer value randomly in the range [0, 100]. 
If there is one constant involved in a clause, the value of the constant \textit{c} is used.
If there are more than one constants involved in a clause, the minimum value from constants is used as a resultant constant \textit{c}.
After determining the constant value, the range is calculated using Equations~\ref{eq4} and~\ref{eq5}.

For example, consider a constraint \textit{C6} as shown in Listing~\ref{lst:re}.
The constraint contains a pre-condition with a single clause, i.e., \lq self.distance+1500 $>$ minDist+maxDist\rq.
In this clause, \textit{distance} is the attribute, the constant \textit{c} is 1500, and \textit{minDist} and \textit{maxDist} are the variables.
If the value of the scaling factor is set to 1, i.e., $s_f = 1$, the minimum and maximum values are calculated as $ A_{min} = \textcolor{black}{0} $ and $ A_{max} = \textcolor{black}{2 * 1500 * 1 = 3000} $.
Therefore, the range of the attribute (\textit{A}) with $s_f = 1$ is $[0, 3000]$.
If the value of the scaling factor is set to 2, i.e., $s_f = 2$, the minimum and maximum values are calculated as  $ A_{min} = \textcolor{black}{0} $ and $ A_{max} = \textcolor{black}{2 * 1500 * 2 = 6000} $.
Therefore, the range of the attribute (\textit{A}) with $s_f = 2$ is $[0, 6000]$.

\subsubsection{Constraint with Multiple Clauses}\label{mc}
Let $c=\{c_1, c_2, c_3,\dots, c_n\}$ be the collection of constants corresponding to an attribute in \textit{A} used in multiple clauses.
Let $c_{min} = min(c)$ and $c_{max} = max(c)$ be the minimum and maximum values in the constants collection \textit{c}. 
The minimum and maximum values for the range of attribute \textit{A} are calculated using Equations~\ref{eq6} and~\ref{eq7}. These equations present range calculation formulas corresponding to positive, negative, and zero values of \textit{c}.

\begin{equation}\label{eq6}
  A_{min}\mathrm{=}  \left\{ 
  \begin{array}{ l l }
    \textcolor{black}{0} & \quad c\mathrm{>}0 \\
    \textcolor{black}{2 * c_{min}} * s_f & \quad c\mathrm{<}0 \\
    -1 * s_f & \quad c\mathrm{=}0 
  \end{array}
\right.
\end{equation}

\begin{equation}\label{eq7}
  A_{max}\mathrm{=}  \left\{ 
  \begin{array}{ l l }
    \textcolor{black}{2 * c_{max}} * s_f & \quad c\mathrm{>}0 \\
    \textcolor{black}{0} & \quad c\mathrm{<}0 \\
    1 * s_f & \quad c\mathrm{=}0 
  \end{array}
\right.
\end{equation}

For example, consider a constraint \textit{C7} as shown in Listing~\ref{lst:re}.
There are two constant values, i.e., $c=\{100, 5000\}$ for the attribute \textit{flightDistance} in this constraint. 
The minimum and maximum values in the constants collection \textit{c} are $c_{min} = 100$ and $c_{max} = 5000$ respectively.
If the value of the scaling factor is set to 1, i.e., $s_f = 1$, the minimum and maximum values for the range are  $ A_{min} = \textcolor{black}{0} $ and $ A_{max} = \textcolor{black}{2 * 5000 *  = 10000} $.
Therefore, the range of the attribute \textit{flightDistance} is $[0, 10000]$.

\textbf{Case 1: Constants with arithmetic manipulation.}
Let $c=\{c_1, c_2, c_3,\dots, c_n\}$ be the collection of constants corresponding to an attribute in \textit{A} containing arithmetic operators.
First, solve the constants expression according to the precedence of the arithmetic operators.
Let $c_a=\{c_{1_a}, c_{2_a}, c_{3_a},\dots, c_{n_a}\}$ be the collection of constants obtained by solving all arithmetic expressions.
Let $c_{min} = min(c_a)$ and $c_{max} = max(c_a)$ be the minimum and maximum values in the constants collection \textit{$c_a$}. 
The minimum and maximum values for the range of the attribute \textit{A} are calculated using Equations~\ref{eq6} and~\ref{eq7}.

For example, consider a constraint \textit{C8} as shown in Listing~\ref{lst:re}.
The constraint \textit{C8} contains an invariant with two clauses, i.e., \lq self.mission.flightDistance $>$ self.uav.MIN\_RANGE+100 and self.mission.flightDistance $>$ self.uav.MAX\_RANGE-100\rq.
In this clause, \textit{flightDistance} is the attribute (\textit{A}) and there are three different constants, i.e., $c_1 = $MIN\_RANGE, $c_2 = 100$, and $c_3 = $MAX\_RANGE.
Both clauses contain constants with addition and subtraction arithmetic operators (i.e., + and -).
If the value of the constant MIN\_RANGE is 20 and MAX\_RANGE is 1000, the resultant constants after solving both arithmetic expressions are $c_{1_a} = 120 $ and $ c_{2_a} = 900$. 
Therefore, the minimum and maximum values in the constants collection \textit{$c_a$} are $ c_{min} = 120 $ and $ c_{max} = 900$.
If the value of the scaling factor is set to 1, i.e., $s_f = 1$, $ A_{min} = \textcolor{black}{0} $ and $ A_{max} = \textcolor{black}{2 * 900 * 1 = 1800} $.
Therefore, the range of the attribute (\textit{A}) with $s_f = 1$ is $[0, 1800]$.

\textbf{Case 2: Multiple clauses joined with Boolean operators and involve different attributes/variables.}
This case considers constraints with multiple clauses containing attributes/variables on both sides of a relational operator and with all clauses independent of each other.
The heuristic is to consider each clause individually and use the heuristics for a single clause as described in Cases 2, 3, and 4 of Section~\ref{sc}.

For example, consider a constraint \textit{C9} as shown in Listing~\ref{lst:re}.
The constraint consists of a pre-condition with two clauses, i.e., \lq self.distance $>$ self.MAX\_RANGE and maxDist $>$ minDist\rq.
Both clauses contain one attribute \textit{distance}, two different variables (i.e., \textit{maxDist} and \textit{minDist}), and a constant \textit{MAX\_RANGE}.
The range for the first clause is calculated by using Equations~\ref{eq4} and~\ref{eq5}.
If the value of MAX\_RANGE is 10 and the value of the scaling factor is set to 1, the range of attribute (\textit{distance}) with $s_f = 1$ is $[0, 10]$.
In the case of the second clause, i.e., \lq maxDist $>$ minDist\rq, the heuristic described in Case 2 of Section~\ref{sc} can be used to calculate the range for both variables.
If the smallest positive integer value selected at random is 5 and the value of the scaling factor is set to 1, i.e., $s_f = 1$, the range of \textit{maxDist} and \textit{minDist} with $s_f = 1$ is $[0, 10]$.

\textbf{Case 3: Multiple clauses joined with Boolean operators and involve attributes/variables on both sides of a relational operator.}
This case considers constraints with multiple clauses containing attributes/variables on both sides of a relational operator and all or some clauses are dependent\footnote{A clause is dependent on another clause if the same attribute/variable is used in both clauses. }.
The dependent clauses contain some dependent attributes/variables and some dependent attributes/variables.
For all independent attributes/variables, the heuristic is to choose the smallest positive integer value randomly in the range [0, 100]. 
For each dependent attribute/variable, the heuristic is to choose the smallest positive integer value randomly in the range [0, 100]. 
The selected Integer value for each dependent and independent attribute/variable is considered a constant \textit{c}.
Each constant determines the range of all attributes or variables using Equations~\ref{eq4} and~\ref{eq5}.

For example, consider a constraint \textit{C10} as shown in Listing~\ref{lst:re}.
The constraint contains a pre-condition with an independent attribute \textit{distance} and two dependent variables \textit{minDist} and \textit{maxDist}.
If the smallest positive integer value selected at random is 5, i.e., $c = 5$.
If the value of the scaling factor is set to 1, the range of \textit{distance} with $s_f = 1$ is $[0, 10]$.
If the smallest positive integer value selected at random for the dependent variable \textit{minDist} is 10, i.e., $c = 10$.
If the value of the scaling factor is set to 1, the range of \textit{distance} with $s_f = 1$ is $[0, 20]$.
Similarly, if the smallest positive integer value selected at random for dependent variable \textit{maxDist}  is 15, i.e., $c = 15$.
If the value of the scaling factor is set to 1, the range of \textit{distance} with $s_f = 1$ is $[0, 30]$.

\subsection{Using Search Algorithms with CBR and Range Reduction}
Search algorithms commonly start by randomly generating solutions and systematically improving them to find the optimal solution(s)~\cite{harman2008search}. 
In our strategy to solve MC/DC constraints, we change the initial random solutions with CBR and range reduction methods. 
Specifically, we select the closest previous solutions using CBR and random solutions with reduced range property values obtained using range reduction. 
Utilizing solutions provided by CBR or range reduction methods, search algorithms work with their specific search mechanisms to find the solution(s). 
We explain this in the context of the AVM search algorithm, which is used in our experiments. 
For example, consider the MC/DC constraint \textit{C1} in Listing~\ref{lst:2} with constant values {\fontfamily{qcr}\selectfont MAX\_TIME=10} and {\fontfamily{qcr}\selectfont MAX\_RANGE=150} and a solution consisting of variables values {\fontfamily{qcr}\selectfont flightTime=8} and {\fontfamily{qcr}\selectfont flightDistance=152}. 
To solve \textit{C2} using \textit{C1}'s solution which is identified as the closest solution by the CBR method, AVM iteratively increments {\fontfamily{qcr}\selectfont flightTime} and decrements {\fontfamily{qcr}\selectfont flightDistance}. 
After a few iterations the solution for \textit{C2} will be found with the variables values {\fontfamily{qcr}\selectfont flightTime=10} and {\fontfamily{qcr}\selectfont flightDistance=149}. 
Similarly, in case of range reduction to solve \textit{C1}, consider that the reduced range for variables {\fontfamily{qcr}\selectfont flightTime \& flightDistance} is {\fontfamily{qcr}\selectfont [0, 300]}. 
Using this range, we generate random solutions for both variables, e.g., consider a random solution with values  {\fontfamily{qcr}\selectfont flightTime=2} and {\fontfamily{qcr}\selectfont flightDistance=231} within the range. 
To solve \textit{C1}, both {\fontfamily{qcr}\selectfont flightTime} and {\fontfamily{qcr}\selectfont flightDistance} values already satisfy C1 and AVM does not need to change these values. 
To solve \textit{C2}, AVM needs to continuously increment {\fontfamily{qcr}\selectfont flightTime} value (i.e., from {\fontfamily{qcr}\selectfont 2} to {\fontfamily{qcr}\selectfont 10}) and decrement {\fontfamily{qcr}\selectfont flightDistance} value (i.e., from {\fontfamily{qcr}\selectfont 231} to {\fontfamily{qcr}\selectfont 149}) until a solution is found. 
Note that with reduced ranges AVM takes fewer iterations to find a solution compared to the full range for Integer variables.

	\section{Empirical Evaluation}
\label{exp}
In this section, we present the empirical evaluation of our proposed strategy with CBR and range reduction methods. 
The goal of empirical evaluation is to analyze the effectiveness of proposed methods (CBR and range reduction).
For the evaluation, we selected the AVM that solves each constraint individually using the approach by Ali et al.~\cite{ali2013generating}. 
We select AVM because the final MC/DC constraints contain several conjunctions, disjunctions, and negations for which AVM has shown better performance than (1+1) EA, as reported in previous studies by Ali et al.~\cite{ali2013generating,ali2015improving}. 
Moreover, experimental results by Ali et al.~\cite{ali2013generating,ali2015improving} show that local search algorithms (e.g., AVM) perform better as compared to global search algorithms (e.g., GA). 
Therefore, our experiment includes AVM as a search algorithm. 
We use Random Search (\rs) as a comparison baseline. 
We evaluate each method of our strategy using AVM to have a fair comparison.
The original AVM is termed \avmo, CBR with AVM is termed \avmc, and range reduction with AVM is termed \avmr. 
The evaluation is also aimed at comparing the combination of CBR and range reduction with the individual methods and is thus termed \avmrc.
Additionally, we compare our proposed methods (i.e., \avmc, \avmr, and \avmrc) for solving MC/DC constraints with relevant constraint solvers. 
The following are the research questions for our empirical evaluation.

\textbf{RQ1:} What is the performance of proposed methods in solving MC/DC constraints?

\textbf{RQ2:} What is the effect of using combinations of both proposed methods (i.e., \avmc~and \avmr) for solving MC/DC constraints?

\textbf{RQ3:} How do the proposed MC/DC constraint-solving methods compare to the existing constraint solvers?

For RQ1, we primarily evaluate \avmc, \avmr, and \avmo~in terms of success rates. 
Whenever a search algorithm solves a constraint, it is considered a success, and its success rate is increased. 
A 100\% success rate of an algorithm corresponding to a constraint means that the algorithm solves the constraint in every run. 
In some cases, when two algorithms achieve the same success rate, comparing the algorithms based on only success rates may not be sufficient to analyze which algorithm is better. 
Therefore, we used iterations count as another measure. 
The number of iterations a search algorithm takes in solving a constraint is termed an iteration count.  
We also compare algorithms based on the time to solve OCL constraints.
Regarding RQ2, we are interested in analyzing the effect of combining \avmc~and \avmr~(i.e., \avmrc) in solving MC/DC constraints. 
For this analysis, we also use success rates, iteration counts, and time as the evaluation metrics. 
In RQ3, we compare our proposed MC/DC constraint-solving methods (\avmc, \avmr, and \avmrc) with relevant constraint solvers, considering the percentage of successfully solved constraints and the average time taken by different solvers.

\begin{table}
	\centering
	\caption{Case studies statistics including the name and type of case study, number of UML classes, total OCL constraints, and the number of OCL constraints used in our experiment}
	\label{tab:cst}
	\begin{tabular}{|l|l|l|l|l|}
		\hline
		\textbf{Name} & \textbf{Type} & \textbf{Classes} & \textbf{OCL} & \textbf{Used}\\
		\hline
		Ground Control Station (\gcs) & Industrial & 38 & 90 & 30 \\  
		EU-Rental (\eur)  & Benchmark & 171 & 243 & 34 \\  
		Royal and Loyal (\rnl)  & Benchmark & 17 & 42 & 10 \\  
		Statemachine Metamodel (\sm) & Benchmark & 34 & 63 & 10 \\  
		Seat Belt System (\sbs) & Artificial & 61 & 25 & 25 \\  	
		Student Registration System (\srs)  & Industrial & 29 & 30 & 20 \\  
		\hline\hline
		\textbf{Total} &  &  & \textbf{493} & \textbf{129} \\  
		\hline
	\end{tabular}
\end{table}

\subsection{Experiment Design and Settings}
For the experiment, we used six cross-domain case studies with different complexity levels. 
Table~\ref{tab:cst} shows statistics for each case study. 
One of the case studies is industrial which represents the software system of a Ground Control Station (\gcs) used to remotely control Unmanned Aerial Vehicles~\cite{sartaj2021automated,sartaj2024automated}. 
Two case studies, EU-Rental~\cite{frias2003eu} and the Royal and Loyal~\cite{warmer2003object,oclrepo} are widely referred to in OCL literature and are considered as benchmark case studies~\cite{khan2019aspectocl}.  
The fourth case study represents the UML meta-model of State Machines~\cite{uml}. 
The fifth case study is an artificial case study (Seat Belt System) that we developed inspired by various industrial case studies published in the literature~\cite{ali2013generating,ali2015improving}. 
The final case study is industrial which represents a Student Registration System (\srs). 

The total number of OCL constraints in all case studies is 493.
Out of 493 OCL constraints, for our experiment, we select constraints with more than one predicate and their outcome is a Boolean value (a requirement for solving a constraint~\cite{ali2013generating}). 
This results in 129 total OCL constraints. 
The OCL constraints used in our experiment have different levels of complexity and contain predicates ranging from two to seven clauses. 
The constraints' complexity depends on the number of clauses, variables, and their conjunctions that are complicated to solve. For instance, for the example constraints given in Listing~\ref{lst:re}, \textit{C2} with one clause and one variable are easy to solve as compared to \textit{C1} with three clauses and three variables. Similarly, the constraint with seven and multiple variables will be more complex to solve.

Due to randomness in the search algorithms, we repeated each algorithm 100 times for all constraints. 
We used iterations (fitness evaluations) as a stopping criterion. 
We set the maximum number of iterations to 2000 (used earlier by Ali et al.~\cite{ali2013generating}). 
The experiment is executed on 15 independent machines including five machines with quad-core 1.2 GHz processor and 1 GB RAM, five machines with core i5 1.7 GHz processor and 8 GB RAM, and five machines with core i7 3.2 GHz processor and 32 GB RAM. 
All algorithms for the same case study are executed on a group of machines with the same specifications.

We analyze experiment results according to the guidelines presented by Arcuri et al.~\cite{arcuri2011practical}. 
Our primary goal is to analyze search algorithms based on success rates. 
This leads to the dichotomous type of data. 
For this purpose, we use Fisher's exact test to analyze the statistical significance and calculate the Odds Ratios to check which algorithm is better. 
In some cases, two search algorithms may have the same success rate and the comparison results are not statistically significant. 
Therefore, we also used iterations count as a measure. 
In the case algorithms are compared based on iterations count, we use the Wilcoxon test for statistical significance and Vargha-Delaney \^{A}${}_{12}$ measure~\cite{vargha2000critique} to analyze which one is better.  
We set the statistical significance level to 0.05 (i.e., $\alpha=0.05$) which is an accepted practice in the domain~\cite{ali2013generating,ali2015improving}.

For the comparison of our methods with existing constraint solvers (RQ3), we conducted an experiment using all six case studies, comprising a total of 129 OCL constraints. We selected two constraint solvers for this comparison: a CSP-based solver, \umltocsp~\cite{cabot2014verification}, and a SMT+Search-based solver, \pledge~\cite{soltana2020practical}. We chose these two constraint solvers to ensure a fair comparison. Both constraint solvers take UML class diagrams and OCL constraints as input, without requiring the transformation of OCL constraints into an intermediate formalism like Alloy, which would introduce additional overhead for industrial constraints~\cite{ali2013generating}. The constraint solver we used operates similarly, making it a suitable comparison. 
\textcolor{black}{
In addition, our selection was guided by the intent to utilize representative approaches well-regarded in the field. 
\umltocsp, for instance, is a well-established approach that has been extensively studied in the literature~\cite{soltana2020practical}. Similarly, \pledge~represents a state-of-the-art hybrid constraint-solving approach. 
This selection ensures good representative approaches for comparison. 
}
We executed this experiment on one machine with a 3.2 GHz processor, 32 GB RAM, and a Windows 10 operating system.

\subsection{Discussion on Results}
\begin{figure*}
	\subfigure[\gcs]{\includegraphics[width=9cm,height=6.5cm]{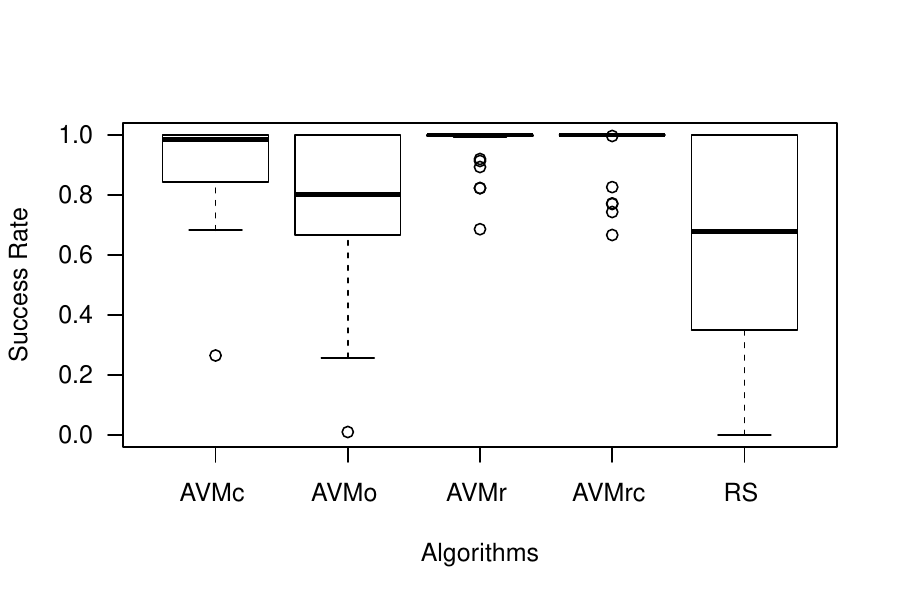}\label{fig:srgcs}}
	\subfigure[\eur]{\includegraphics[width=9cm,height=6.5cm]{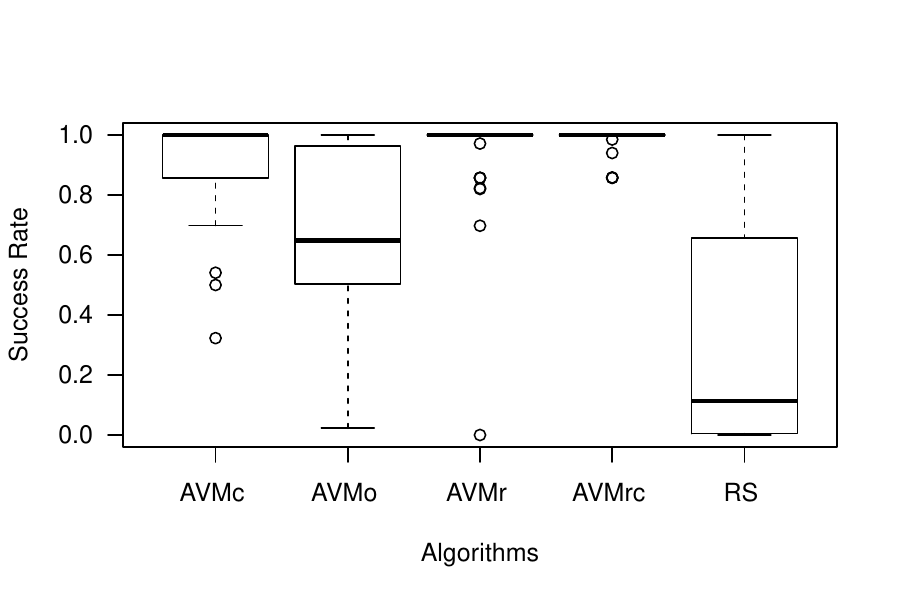}\label{fig:sreur}}
	\subfigure[\rnl]{\includegraphics[width=9cm,height=6.5cm]{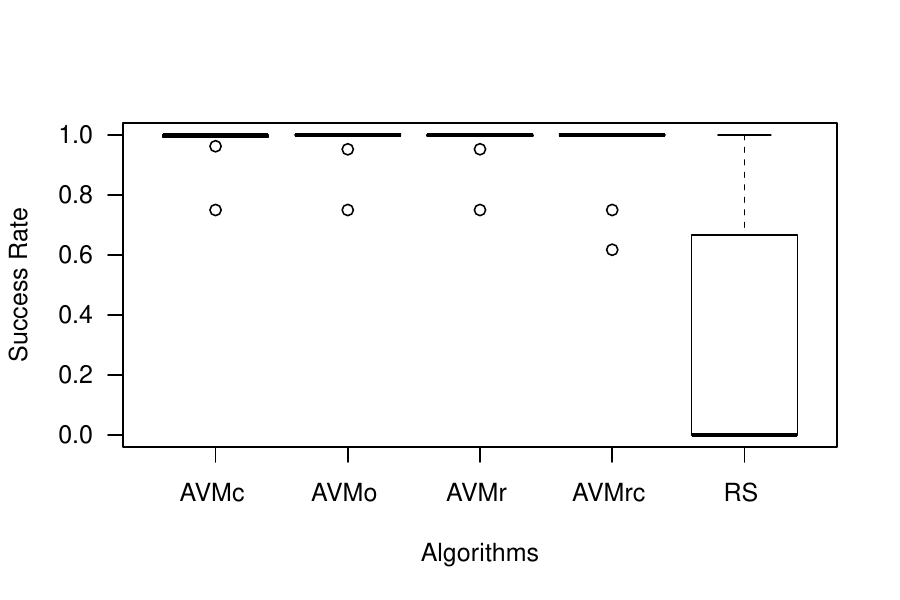}\label{fig:srrnl}}
	\subfigure[\sm]{\includegraphics[width=9cm,height=6.5cm]{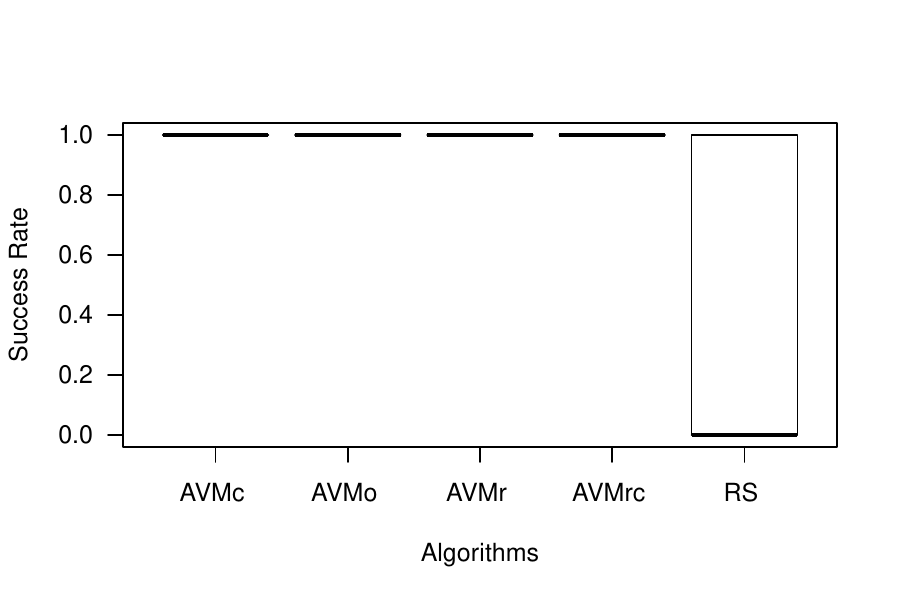}\label{fig:srsm}}
	\subfigure[\sbs]{\includegraphics[width=9cm,height=6.5cm]{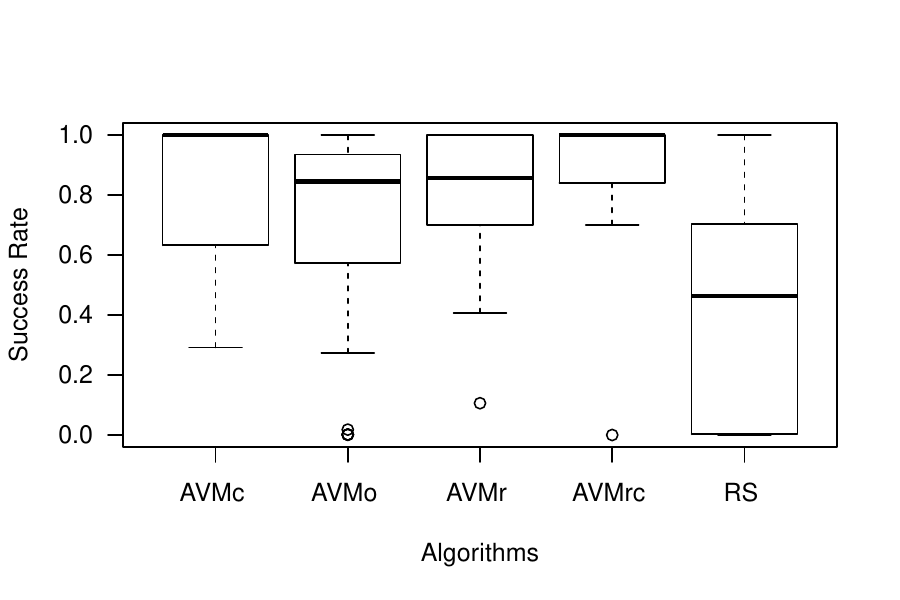}\label{fig:srsbs}}
	\subfigure[\srs]{\includegraphics[width=9cm,height=6.5cm]{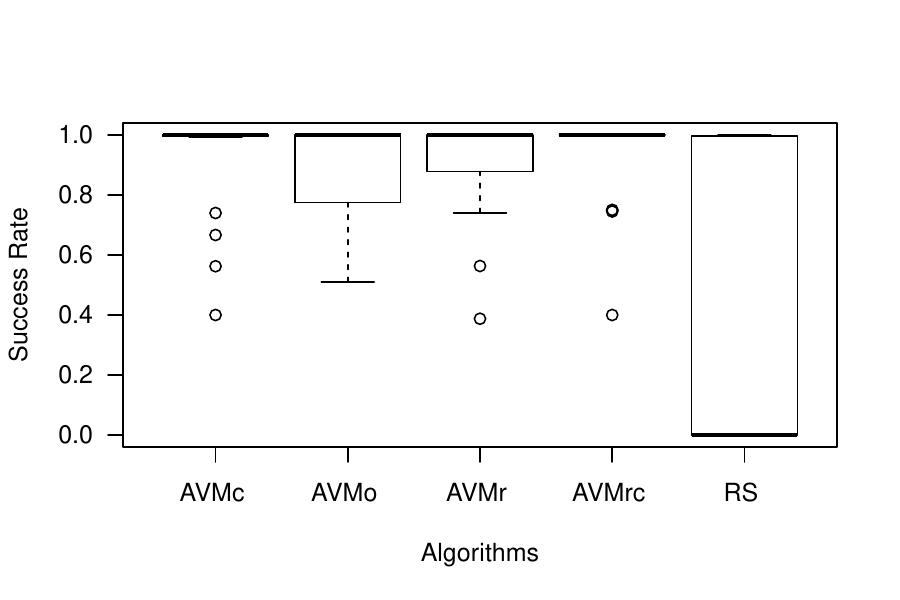}\label{fig:srsrs}}
	\caption{Boxplots showing the median success rates achieved by \avmo, \avmc, \avmr, \avmrc, and \rs~for each case study.}\label{fig:srcomp}
\end{figure*}

\begin{figure*}
	\subfigure[\gcs]{\includegraphics[width=9cm,height=6.5cm]{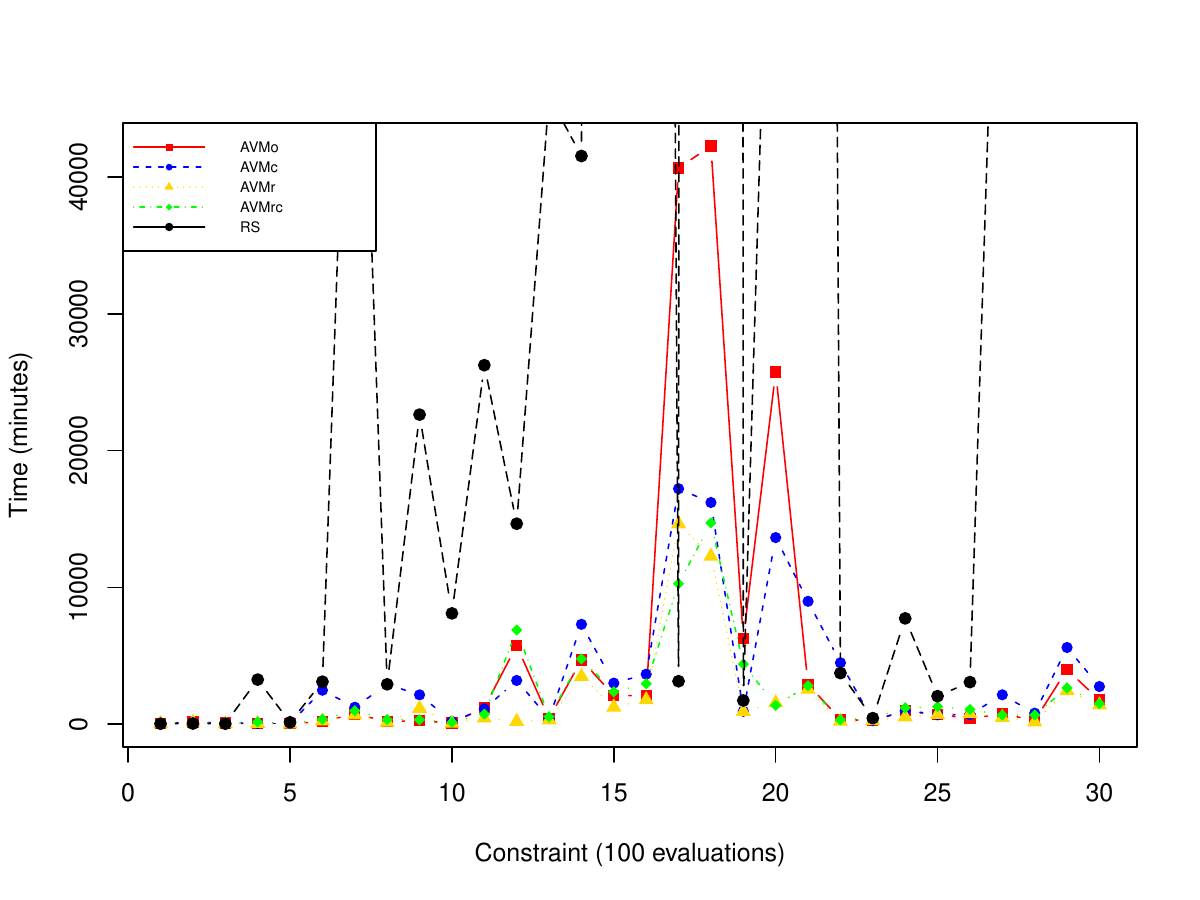}\label{fig:gcst}}
	\subfigure[\eur]{\includegraphics[width=9cm,height=6.5cm]{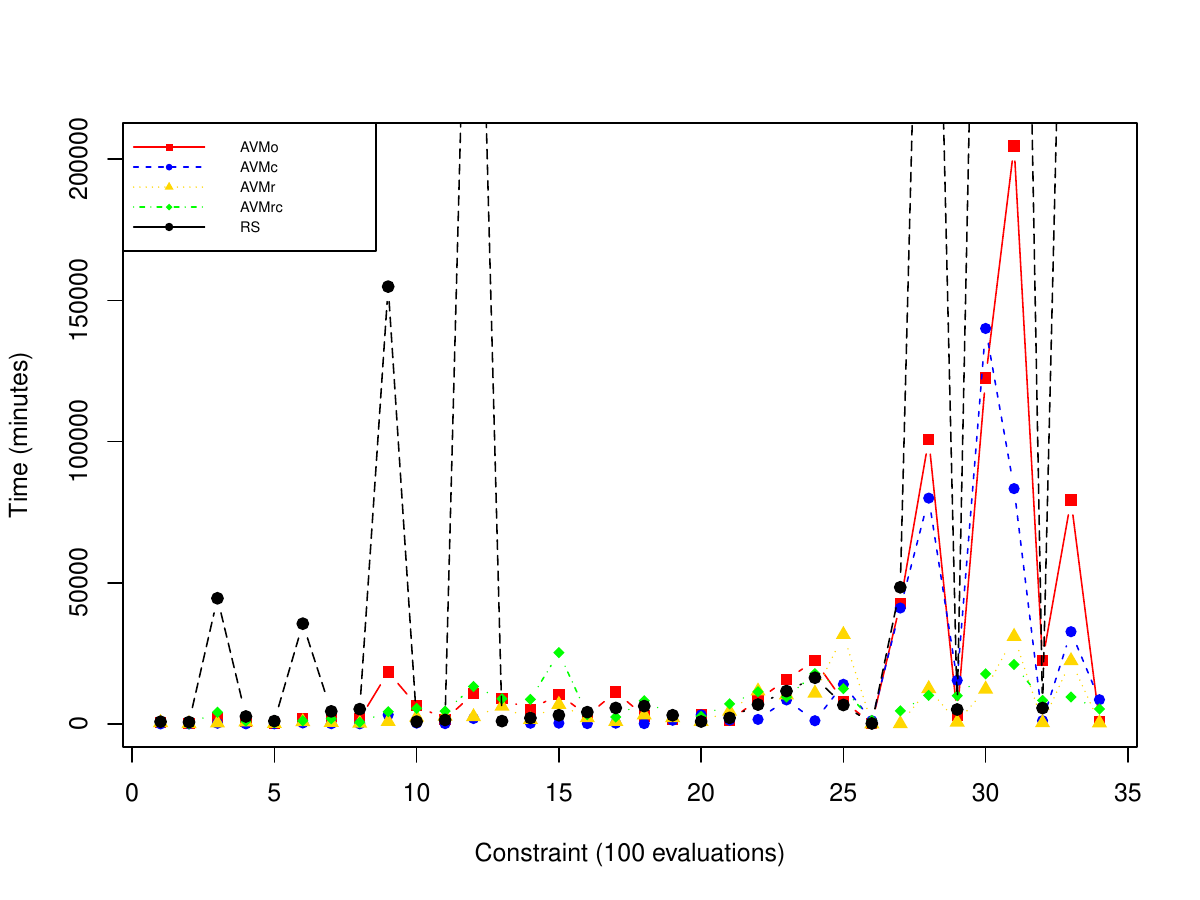}\label{fig:eurt}}	
	\subfigure[\rnl]{\includegraphics[width=9cm,height=6.5cm]{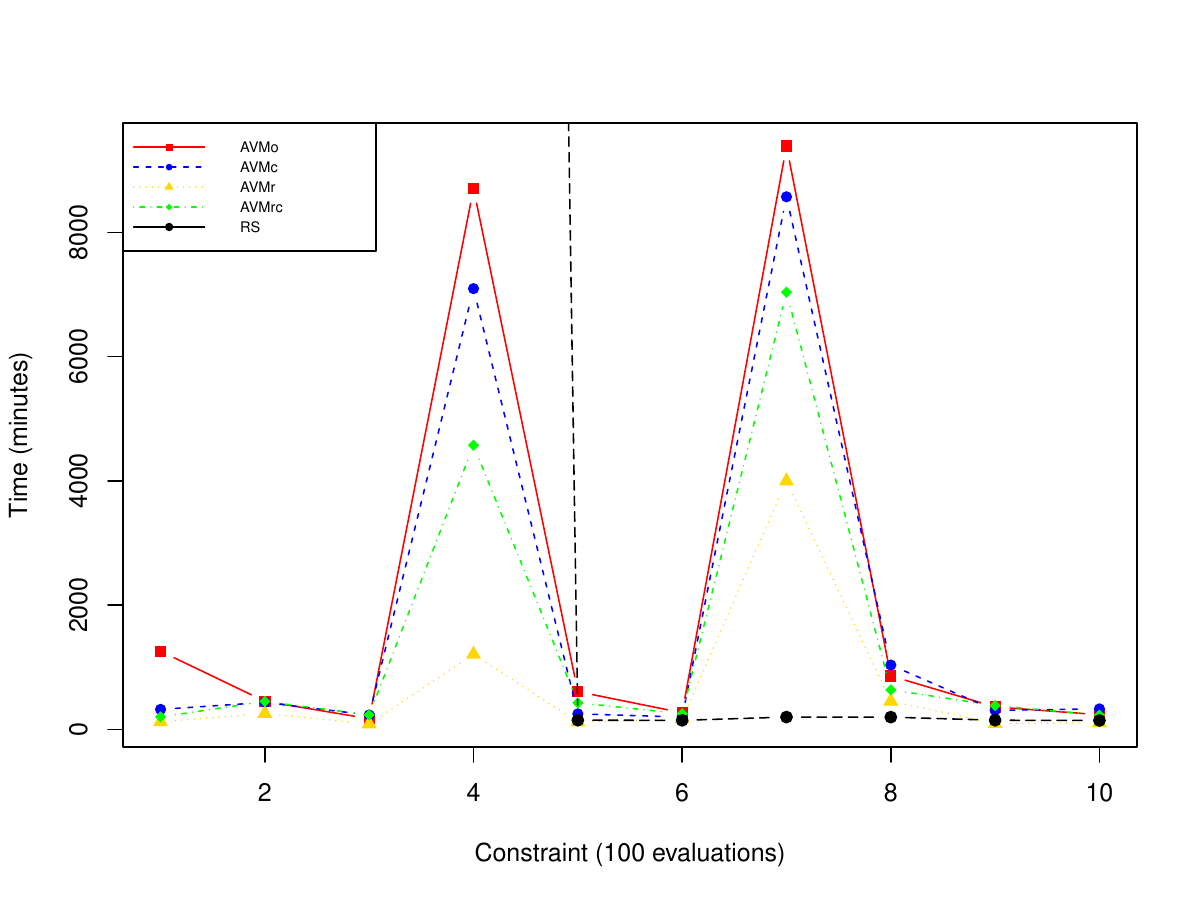}\label{fig:rnlt}}
	\subfigure[\sm]{\includegraphics[width=9cm,height=6.5cm]{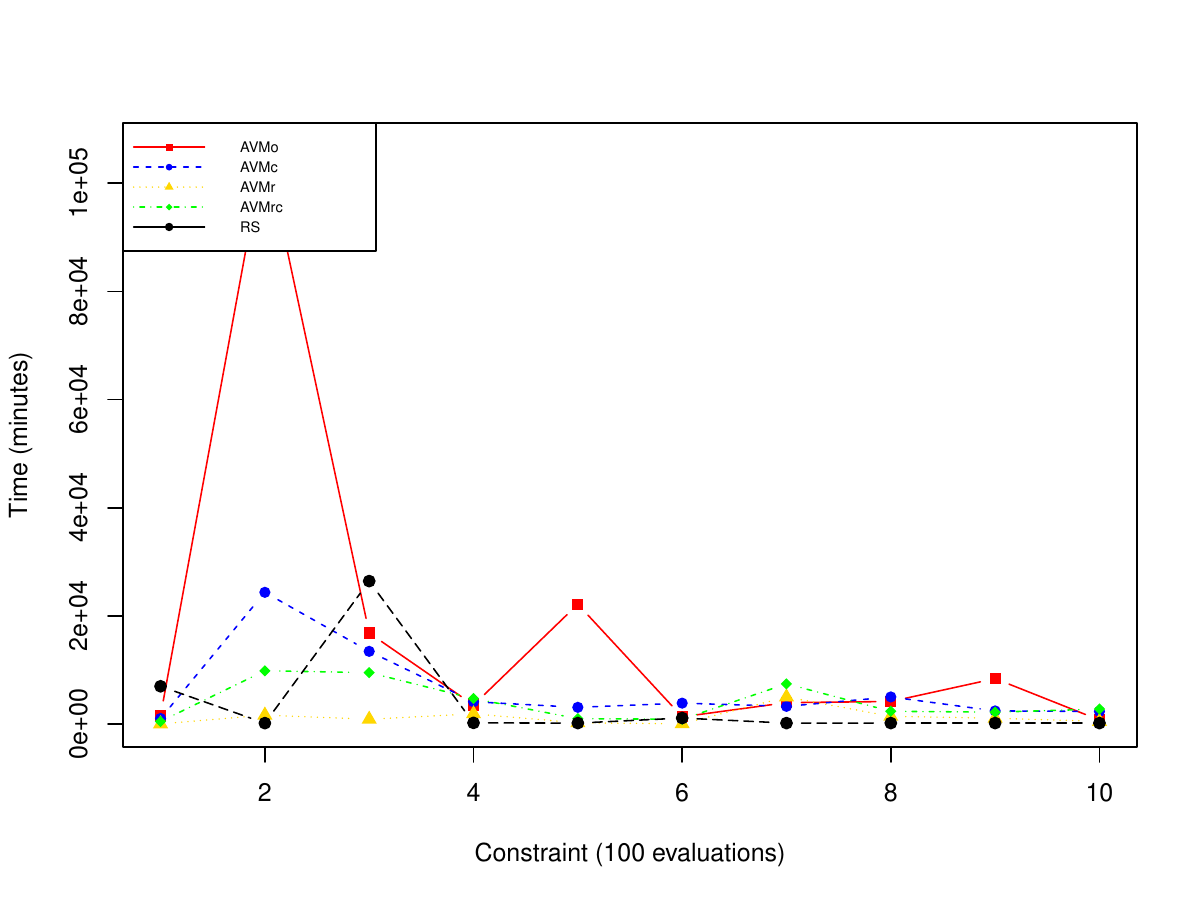}\label{fig:smt}}
	\subfigure[\sbs]{\includegraphics[width=9cm,height=6.5cm]{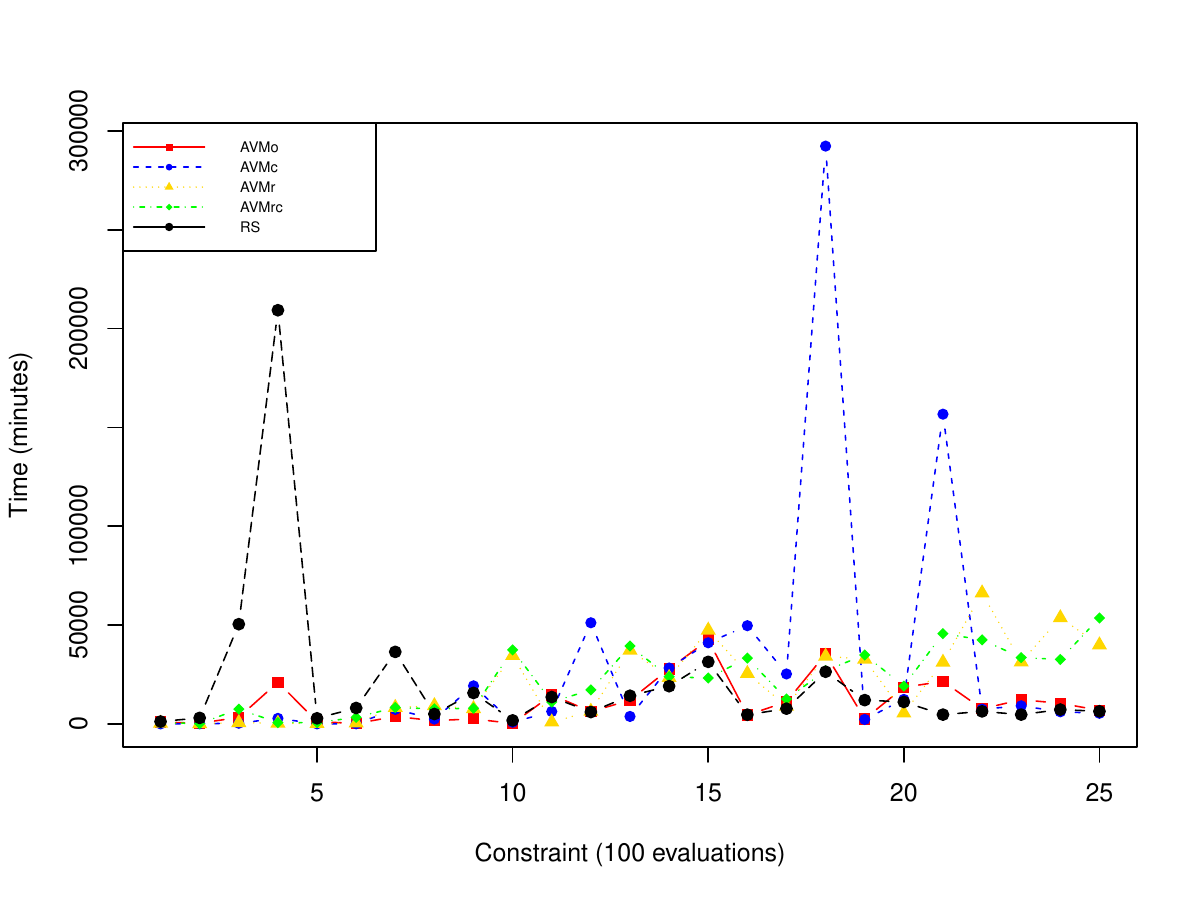}\label{fig:sbst}}
	\subfigure[\srs]{\includegraphics[width=9cm,height=6.5cm]{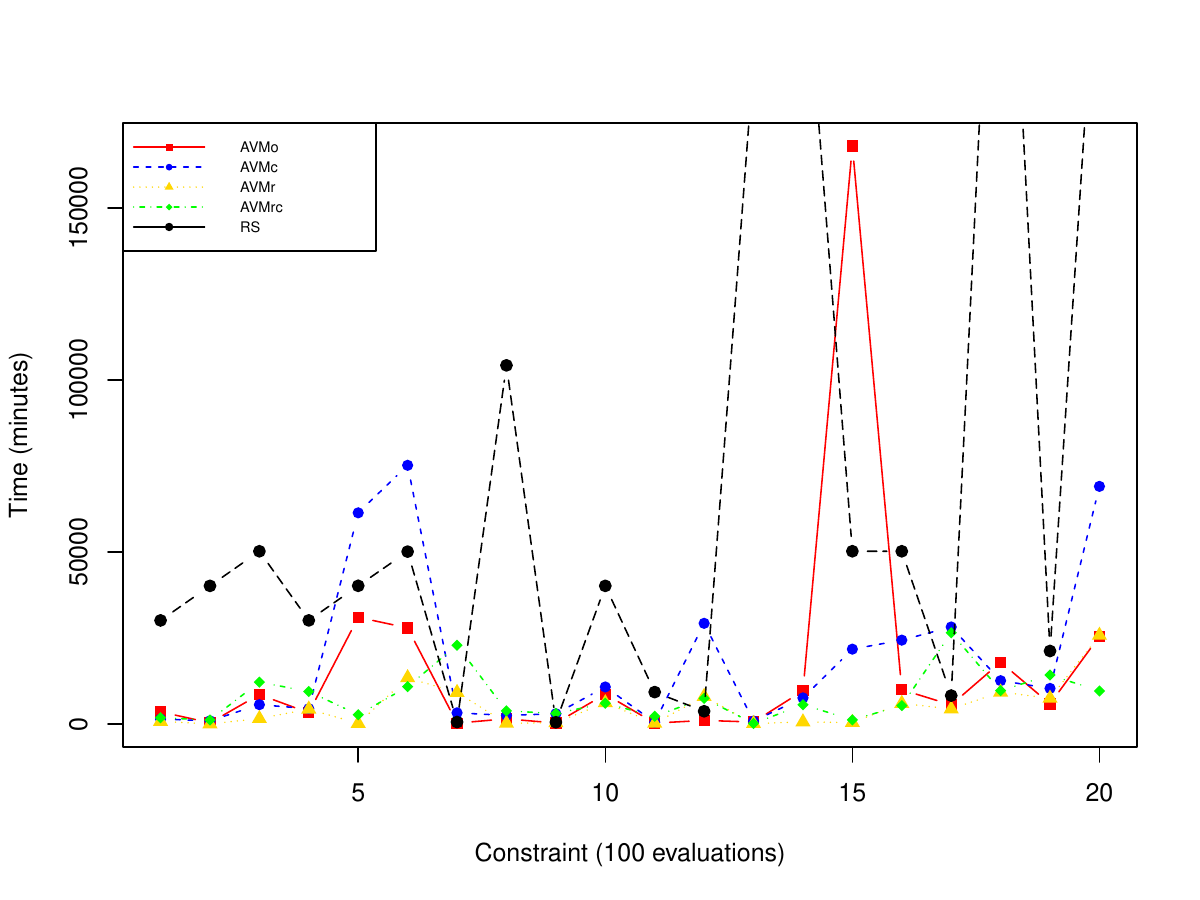}\label{fig:srst}}
	\caption{Line graphs showing the time taken by \avmo, \avmc, \avmr, \avmrc, and \rs~to solve MC/DC constraints for each case study across 100 evaluations.}\label{timecomp}
\end{figure*}

Figure~\ref{fig:srcomp} shows the success rates achieved by \avmr, \avmc, \avmrc, \avmo, and \rs~for all case studies. 
Table~\ref{tab:all} shows the results of the Fisher Exact test and the Wilcoxon test for statistical significance and the results of the effect size calculated using Odds Ratios and Vargha-Delaney's \^{A}${}_{12}$ measure. 
For the comparison based on time, Figure~\ref{timecomp} shows the time taken for each OCL constraint during 100 evaluations. 
Table~\ref{tab:time} shows the results of the Wilcoxon test for statistical significance and the results of the effect size calculated using Vargha-Delaney's \^{A}${}_{12}$ measure. 
The columns in both tables represent a comparison between two algorithms based on statistical significance and effect size corresponding to each case study and its total constraints. 
The values presented in both tables (i.e., Table~\ref{tab:all} and Table~\ref{tab:time}) refer to the number of times p-value$<$0.05 for the results of Fisher Exact and the Wilcoxon test. 
Similarly, for the effect size measures (Odds Ratios and Vargha-Delaney's \^{A}${}_{12}$), the tables show the number of times (including percentage fraction) an algorithm is better than the other one, i.e., Algorithm\_1 outperforms ($>$) Algorithm\_2.
We show aggregated results due to a large number of comparisons among algorithms corresponding to all case studies. Table~\ref{tab:rq3} presents results of comparison among \avmr, \avmc, \avmrc, \umltocsp, and \pledge~based on the percentage of solved MC/DC constraints and the average time.

\begin{table}
	\centering
	\caption{The results of algorithms comparison based on success rates and iteration counts}
	\label{tab:all}
	\small
	\begin{tabular}{|l|l|l|l|l|l|l|l|}
		\hline
		\multicolumn{2}{|l|}{\textbf{Comparison}} & \textbf{\gcs~(30)} & \textbf{\eur~(34)} & \textbf{\rnl~(10)} & \textbf{\sm~(10)} & \textbf{\sbs~(25)} & \textbf{\srs~(20)}\\    
		\hline
		\multirow{2}{*}{\avmc~vs \avmo} &SRC (\# p-values  $<\alpha$)&24&28&8&10&24&15\\  \cline{2-8}
		&ES (\avmc~$>$ \avmo)&22 (73\%)&26 (87\%)&8 (80\%)&3 (30\%)&19 (76\%)&3 (15\%)\\  \cline{1-8}
		\multirow{2}{*}{\avmc~vs \avmr} &SRC (\# p-values $<$ $\alpha$)&25&30&8&9&23&18\\  \cline{2-8}
		&ES (\avmc~$>$ \avmr)&15 (50\%)&19 (56\%)&8 (80\%)&6 (60\%)&14 (56\%)&10 (50\%)\\  \cline{1-8}
		\multirow{2}{*}{\avmc~vs \avmrc} &SRC (\# p-values $<$ $\alpha$)&27&28&9&9&25&17\\  \cline{2-8}
		&ES (\avmc~$>$ \avmrc)&15 (50\%)&16 (47\%)&9 (90\%)&6 (60\textcolor{black}{\%})&14 (56\%)&10 (50\%)\\  \cline{1-8}

		\multirow{2}{*}{\avmr~vs \avmo}  &SRC (\# p-values $<$ $\alpha$)&29&31&8&7&25&17\\  \cline{2-8}
		&ES (\avmr~$>$ \avmo)&20 (67\%)&28 (82\%)&5 (50\%)&1 (10\%)&14 (56\%)&6 (30\%)\\  \cline{1-8}
		\multirow{2}{*}{\avmrc~vs \avmo} &SRC (\# p-values $<$ $\alpha$)&27&31&9&9&25&18\\  \cline{2-8}
		&ES (\avmrc~$>$ \avmo)&20 (67\%)&30 (88\%)&3 (30\%)&2 (20\%)&17 (87\%)&6 (30\%)\\  \cline{1-8}
		\multirow{2}{*}{\avmr~vs \avmrc} &SRC (\# p-values $<$ $\alpha$)&28&26&8&10&22&16\\  \cline{2-8} 
		&ES (\avmr~$>$ \avmrc)&20 (67\%)&18 (53\%)&7 (70\%)&7 (70\%)&14 (56\%)&13 (65\%)\\  \cline{1-8}
		\multirow{2}{*}{\avmo~vs \rs}   &SRC (\# p-values $<$ $\alpha$)&28&32&10&7&20&16\\    \cline{2-8}
		&ES (\avmo~$>$ \rs)&18 (60\%)&31 (91\%)&9 (90\%)&7 (70\%)&19 (76\%)&15 (75\%)\\  \cline{1-8}
		\multirow{2}{*}{\avmc~vs \rs}  &SRC (\# p-values $<$ $\alpha$)&27&32&10&9&23&17\\  \cline{2-8}
		&ES (\avmc~$>$ \rs)&21 (70\%)&32 (94\%)&9 (90\%)&9 (90\%)&22 (88\%)&15 (75\%)\\  \cline{1-8}
		\multirow{2}{*}{\avmr~vs \rs} &SRC (\# p-values $<$ $\alpha$)&25&32&10&9&25&19\\   \cline{2-8}
		&ES (\avmr~$>$ \rs)&21 (70\textcolor{black}{\%})&32 (94\%)&9 (90\%)&9 (90\%)&22 (88\%)&15 (75\%)\\  \cline{1-8}
		\multirow{2}{*}{\avmrc~vs \rs} &SRC (\# p-values $<$ $\alpha$)&27&32&10&9&25&19\\ \cline{2-8} 
		&ES (\avmrc~$>$ \rs)&23 (76\textcolor{black}{\%})&32 (94\%)&9 (90\%)&9 (90\%)&21 (84\%)&16 (80\%)\\ 
		\hline
        \multicolumn{8}{l}{*Significant results count (SRC) shows the number of times $p-values<\alpha$ and the effect size (ES) count shows the}\\
		\multicolumn{8}{l}{number of instances (and by what percentage) an algorithm outperforms the other, i.e., Algorithm\_1 $>$ Algorithm\_2.} \\
	\end{tabular}
\end{table}

\begin{table}
	\centering
	\caption{The results of algorithms comparison based on the time to solve MC/DC constraints}
	\label{tab:time}
	\small
	\begin{tabular}{|l|l|l|l|l|l|l|l|}
		\hline
		\multicolumn{2}{|l|}{\textbf{Comparison}} & \textbf{\gcs~(30)} & \textbf{\eur~(34)} & \textbf{\rnl~(10)} & \textbf{\sm~(10)} & \textbf{\sbs~(25)} & \textbf{\srs~(20)}\\  
		\hline
		\multirow{2}{*}{\avmc~vs \avmo} &SRC (\# p-values $<$ $\alpha$)&25&25&8&7&21&19\\  \cline{2-8}
		&ES (\avmc~$>$ \avmo)&15 (50\%)&9 (26\%)&3 (30\%)&3 (30\%)&7 (28\%)&15 (75\%)\\  \cline{1-8}
		\multirow{2}{*}{\avmc~vs \avmr} &SRC (\# p-values $<$ $\alpha$)&26&28&9&10&24&16\\  \cline{2-8}
		&ES (\avmc~$>$ \avmr)&23 (77\%)&6 (18\%)&9 (90\%)&9 (90\%)&3 (12\%)&16 (80\%)\\  \cline{1-8}
		\multirow{2}{*}{\avmc~vs \avmrc} &SRC (\# p-values $<$ $\alpha$)&26&28&7&8&24&19\\  \cline{2-8}
		&ES (\avmc~$>$ \avmrc)&14 (47\%)&2 (5\%)&4 (40\%)&5 (50\%)&2 (8\%)&10 (50\%)\\  \cline{1-8}

		\multirow{2}{*}{\avmr~vs \avmo}  &SRC (\# p-values $<$ $\alpha$)&27&29&10&10&23&18\\  \cline{2-8}
		&ES (\avmr~$>$ \avmo)&0 (0\%)&19 (56\%)&0 (0\%)&1 (10\%)&20 (80\%)&6 (30\%)\\  \cline{1-8}
		\multirow{2}{*}{\avmrc~vs \avmo} &SRC (\# p-values $<$ $\alpha$)&26&28&7&10&23&17\\  \cline{2-8}
		&ES (\avmrc~$>$ \avmo)&20 (67\%)&24 (71\%)&1 (10\%)&3 (30\%)&21 (84\%)&10 (50\%)\\  \cline{1-8}
		\multirow{2}{*}{\avmr~vs \avmrc} &SRC (\# p-values $<$ $\alpha$)&28&30&10&10&24&18\\  \cline{2-8} 
		&ES (\avmr~$>$ \avmrc)&0 (0\%)&3 (9\%)&0 (0\%)&0 (0\%)&2 (8\%)&4 (10\%)\\  \cline{1-8}
		\multirow{2}{*}{\avmo~vs \rs}   &SRC (\# p-values $<$ $\alpha$)&29&26&10&10&20&19\\    \cline{2-8}
		&ES (\avmo~$>$ \rs)&7 (23\%)&22 (65\%)&8 (80\%)&8 (80\%)&16 (64\%)&1 (5\%)\\  \cline{1-8}
		\multirow{2}{*}{\avmc~vs \rs}   &SRC (\# p-values $<$ $\alpha$)&28&28&9&10&23&20\\  \cline{2-8}
		&ES (\avmc~$>$ \rs)&6 (20\%)&18 (53\%)&6 (60\%)&8 (80\%)&15 (60\%)&5 (25\%)\\  \cline{1-8}
		\multirow{2}{*}{\avmr~vs \rs}   &SRC (\# p-values $<$ $\alpha$)&28&28&10&9&25&19\\   \cline{2-8}
		&ES (\avmr~$>$ \rs)&3 (10\%)&21 (62\%)&3 (30\%)&6 (60\%)&22 (88\%)&0 (0\%)\\  \cline{1-8}
		\multirow{2}{*}{\avmrc~vs \rs}  &SRC (\# p-values $<$ $\alpha$)&28&28&9&10&24&19\\ \cline{2-8} 
		&ES (\avmrc~$>$ \rs)&9 (30\%)&25 (74\%)&6 (60\%)&7 (70\%)&22 (88\%)&4 (20\%)\\ 
		\hline
        \multicolumn{8}{l}{*Significant results count (SRC) shows the number of times $p-values<\alpha$ and the effect size (ES) count shows the}\\
		\multicolumn{8}{l}{number of instances (and by what percentage) an algorithm outperforms the other, i.e., Algorithm\_1 $>$ Algorithm\_2.} \\
	\end{tabular}
\end{table}

\subsubsection{RQ1. Performance comparison among \avmr, \avmc, and \avmo}
To compare algorithms based on success rates, the box plots in Figure~\ref{fig:srcomp} show that \avmr~achieved a median success rate of $\approx$100\% for five case studies (i.e., \gcs, \eur, \rnl, \sm, and \srs). 
Similarly, \avmc~achieved $\approx$100\% median success rate for five case studies (i.e., \eur, \rnl, \sm, \sbs, and \srs). 
\avmo~achieved a median success rate of $\approx$100\% for three case studies (i.e., \rnl, \sm, and \srs). 
Whereas \rs~cannot achieve a 100\% median success rate for any of the case studies.

For the statistical comparison between \avmr~and \avmc, Table~\ref{tab:all} shows that \avmc~outperforms \avmr~more than 50\% of the time for all six case studies. 
In the case of comparison between \avmr~and \avmo, \avmr~is better than \avmo~for $\approx$60\% of the time for \gcs, \eur, \rnl, and \sbs~case studies. 
While comparing \avmc~and \avmo, \avmc~is better than \avmo~for more than $\approx$60\% of the constraints for \gcs, \eur, \rnl, and \sbs~case studies. 
Whereas for two case studies, i.e., \sm~and \srs, \avmo~is better than \avmc~and \avmr~for  $\approx$60\% of the cases. 
If we compare \avmr, \avmc, and \avmo~with \rs, we can notice that all three algorithms outperform \rs~for more than $\approx$80\% of the cases for all six case studies. 

Comparing \avmr~and \avmc~based on time, from Figure~\ref{timecomp} and Table~\ref{tab:time}, it is observed that both \avmr~and \avmc~take approximately an equal amount of time on average. 
For \gcs, \rnl, \sm, and \srs~case studies, \avmc~is significantly better than \avmr. 
For the comparison between \avmr~and \avmo, it is observed that \avmo~is better for  $\approx$60\% of the time.  
However, for \eur, \sbs, and \srs~case studies, \avmr~is better than \avmo~for $\approx$60\% of the constraints. 
In the case of the comparison between \avmc~and \avmo, \avmc~outperforms \avmo~for \gcs, \sm, and \srs~in $\approx$60\% of the constraints. 
For the three case studies, i.e., \eur, \rnl, and \sbs, \avmo~outperforms \avmc~in $\approx$60\% of the cases. 
In the case of comparing \avmr, \avmc, and \avmo~with \rs, we can analyze that all three algorithms outperform \rs~for $\approx$50\% of the cases especially for \eur, \rnl, \sm, and \sbs. 

\begin{tcolorbox}[colback=black!1!white]
	\textbf{Answer to RQ1:} Both \avmr~and \avmc~outperform \avmo~based on success rates and iteration counts. However, \avmo~is better than \avmr~and \avmc~in terms of time. 
\end{tcolorbox}

\subsubsection{RQ2. Analysis of the combination of \avmr~and \avmc}
From the box plots shown in Figure~\ref{fig:srcomp}, it is clear that \avmrc~achieved a median success rate of $\approx$100\% for all six case studies. 
Whereas individually the \avmr~and \avmc~achieved $\approx$100\% median success rate for five case studies. 
Moreover, \avmo~achieved a median success rate of $\approx$100\% for only three case studies. 

The results of statistical comparison based on success rates and iteration counts show that \avmrc~outperforms \avmc~and \avmo~for $\approx$60\% of the cases. 
Specifically, \avmrc~outperforms \avmc~and \avmo~for \gcs, \eur, \sbs, and \srs~case studies.
Moreover, for the comparison between \avmrc~and \avmr, the \avmr~performs better than \avmrc~for all six case studies. 
The comparison between \avmrc~and \rs~shows that \avmrc~is better than \rs~for more than $\approx$85\% of the cases for all six case studies. 

For the comparison based on time, the graph in Figure~\ref{timecomp} shows that \avmrc~takes less time as compared to \avmr, \avmc, and \avmo. 
The statistical comparison results given in Table~\ref{tab:time} show that \avmrc~performs better than \avmo~for more than $\approx$60\% of the constraints specifically for \gcs, \eur, \sbs, and \srs~case studies. 
While comparing \avmrc~and \avmc, it is noticed that \avmrc~is better than \avmc~for more than $\approx$70\% of the cases specifically for \eur, \sbs, and \srs~case studies. 
Whereas for the \gcs, \rnl, and \sm~case studies, \avmc~is better than \avmrc~for $\approx$60\% of the cases. 
In the comparison between \avmrc~and \avmr, it is observed that \avmrc~outperforms \avmr~for more than $\approx$90\% of the time for all six case studies. 
Finally, for the comparison between \avmrc~and \rs, it is clear that \avmrc~performs better than \rs~for $\approx$60\% of the cases specifically for \eur, \rnl, \sm, and \sbs~case studies.

\begin{tcolorbox}[colback=black!1!white]
	\textbf{Answer to RQ2:} The combination of \avmr~and \avmc~(i.e., \avmrc) outperforms individual \avmr, \avmc, and \avmo~based on success rates, iteration counts, and time to solve OCL constraints.
\end{tcolorbox}

\begin{table}
	\centering
	\caption{The comparison results among \avmc, \avmr, \avmrc, \umltocsp, and \pledge~based on the percentage of MC/DC constraints solved and average time (in seconds)}
	\label{tab:rq3}
    
	\begin{tabular}{|l|l|l|l|l|l|l|l|}
		\hline
		\multicolumn{2}{|l|}{\textbf{Approach}} & \textbf{\gcs~(30)} & \textbf{\eur~(34)} & \textbf{\rnl~(10)} & \textbf{\sm~(10)} & \textbf{\sbs~(25)} & \textbf{\srs~(20)}\\    
		\hline
		\multirow{2}{*}{\textbf{\avmc}} &Solved &90.8\%&90.9\%&97.1\%&100\%&82.3\%&91.8\%\\  \cline{2-8}
		&Time (s)&57.31&44.01&5.38&17.57&1.53&24.82\\  \cline{1-8}
		\multirow{2}{*}{\textbf{\avmr}} &Solved &96.4\%&96.1\%&95.3\%&100\%&84.7\%&92.1\%\\  \cline{2-8}
		&Time (s)&31.12&129.34&2.05&0.94&7.84&13.24\\  \cline{1-8}
		\multirow{2}{*}{\textbf{\avmrc}} &Solved &95.9\%&98.5\%&93.7\%&100\%&88.4\%&93.2\%\\  \cline{2-8}
		&Time (s)&86.30&507.82&3.37&8.42&56.10&31.05\\  \cline{1-8}
		\multirow{2}{*}{\textbf{\umltocsp}}  &Solved &29.5\%&31.1\%&43.7\%&39.5\%&18.2\%&25.4\%\\  \cline{2-8}
		&Time (s)&2.81&13.62&1.30&1.46&2.59&2.16\\  \cline{1-8}
		\multirow{2}{*}{\textbf{\pledge}} &Solved &46.3\%&58.5\%&82.5\%&79.9\%&34.2\%&41.1\%\\  \cline{2-8}
		&Time (s)&73.55&274.80&3.12&4.08&24.76&18.95\\  
		\hline
	\end{tabular}
\end{table}

\subsubsection{RQ3. Comparison of \avmc, \avmr, and \avmrc~with \umltocsp~and \pledge}

The comparison results in Table~\ref{tab:rq3} show that \avmc~solved more than 90\% MC/DC constraints for \gcs, \eur, \rnl, and \srs. 
\avmc~solved all constraints for \sm~and $\approx$82\% for \sbs. 
\avmr~solved MC/DC constraints $\approx$96\% for \gcs, $\approx$96\% for \eur, $\approx$95\% for \rnl, $\approx$100\% for \sm, $\approx$84\% for \sbs~and $\approx$92\% for \srs. 
\avmrc~solved MC/DC constraints $\approx$96\% for \gcs, $\approx$98\% for \eur, $\approx$94\% for \rnl, $\approx$100\% for \sm, $\approx$88\% for \sbs~and $\approx$93\% for \srs. 
\umltocsp~was able to solve $\approx$30\% MC/DC constraints for \gcs~and \eur, $\approx$44\% for \rnl, $\approx$40\% for \sm, $\approx$18\% for \sbs~and $\approx$25\% for \srs. 
\pledge~was able to solve MC/DC constraints $\approx$46\% for \gcs, $\approx$59\% \eur, $\approx$83\% for \rnl, $\approx$80\% for \sm, $\approx$34\% for \sbs~and $\approx$41\% for \srs.

Analyzing the average time taken by each constraint-solving method, it is evident that \avmr~required less time compared to \avmc~and \avmrc~across five case studies. \avmr~took more time than \avmc~only for the \eur~case study.
Overall, \umltocsp~took less time compared to all constraint-solving methods. Despite its speed, \umltocsp~was unable to solve a majority of MC/DC constraints, especially for industrial case studies (\gcs~and \srs). 
\pledge~showed faster execution time compared to \avmc~and \avmrc~for the \rnl~and \sm~case studies. 
Additionally, compared to \avmrc, \pledge~consistently took less time for each case study. 
It is important to note that \avmc, \avmr, and \avmrc~outperformed \umltocsp~and \pledge~when considering the percentage of solved MC/DC constraints.

After a thorough analysis of \umltocsp~and \pledge~results, we observed that the reason for unsolved MC/DC constraints is \textcolor{black}{conflicting clauses}. 
Given that these solvers are not specifically designed for solving MC/DC constraints and handling conflicting constraints, a substantial number of constraints remain unsolved.
For instance, \pledge~operates by \textcolor{black}{taking MC/DC constraints and constructing a syntax tree of these constraints. 
It then determines which clauses to solve using SMT and which ones with search. }
During the constraint-solving process, if the clauses to solve using SMT turn out to be conflicting, SMT marks the constraint as unsatisfiable, resulting in unsolved constraints. 
\textcolor{black}{
For example, the MC/DC constraints \textit{C1} and \textit{C2}, as shown in Listing~\ref{lst:2}, contain conflicting clauses. 
These conflicts will be identified as unsatisfiable by the SMT and, as a result, will remain unsolved by \pledge.  
Our strategy starts with an MC/DC constraint and assesses it for potential conflicts. 
If a constraint is conflict-free, our strategy employs CBR or range reduction methods to solve it. 
In scenarios where a conflict occurs and is deemed unsatisfiable, our strategy bypasses the conflicting constraint and proceeds to the next constraint. 
This stepwise constraint-solving process systematically addresses individual constraint conflicts and excludes an MC/DC constraint if it is unsolvable. 
}

\begin{tcolorbox}[colback=black!1!white]
    \textbf{Answer to RQ3:} \avmrc, despite its longer execution time, outperformed in solving a substantial number of MC/DC constraints compared to other approaches. \umltocsp~and \pledge~demonstrated faster execution but struggled to solve the majority of MC/DC constraints from industrial case studies.
\end{tcolorbox}

\subsection{Threats to Validity}
Like any empirical evaluation, our experiment is also subjected to different threats that can affect its validity. 
The possible threats to the validity of our experiment and their mitigation approaches are discussed individually in the following subsections. 

\subsubsection{External Validity Threat}
The threat to external validity is related to the generalization of experiment results. 
To reduce this threat, we selected six case studies from various domains with different sizes and complexities. 
We used 129 OCL constraints from all the case studies and selected constraints that contain predicates with the number of clauses ranging from two to seven. 
The selected constraints were representative of various cases with different complexity levels. 
We compare \avmo~using Ali et al.~\cite{ali2013generating} approach with our proposed strategy implemented in the form of \avmc, \avmr, and \avmrc. 
We also compare the MC/DC constraint-solving methods of our strategy with relevant constraint-solving approaches presented in the literature. 
In our experiment, we tried to reduce the chances of this threat by using a variety of cross-domain case studies and conducting comparisons with existing approaches. 
However, this threat is common in most experiments.

\subsubsection{Construct Validity Threat}
The threat to construct validity occurs when the relationship between cause and effect is undetermined. 
To minimize the chances of this threat, we used the number of iterations (fitness evaluations) as a stopping criterion. 
We set the maximum number of iterations to 2000. 
We also used measures such as success rates, iteration counts, and time that are commonly used to compare various search algorithms.

\subsubsection{Conclusion Validity Threat}
The threat to conclusion validity occurs when treatments can impact the outcome of the experimental results. 
To minimize the effect of this threat, we repeated all OCL constraints 100 times for each treatment. 
We analyzed the statistical significance of experimental results based on the success rates using Fisher's exact test and iteration counts using the Wilcoxon test. 
We also measured the effect size by calculating Odds ratios and using Vargha-Delaney's \^{A}${}_{12}$ measure~\cite{vargha2000critique}. 
We analyzed the outcomes of the comparison with existing constraint solvers based on the percentage of successfully solved constraints and the average time.

\subsubsection{Internal Validity Threat}
The threat to the internal validity of the experiment is associated with the parameter tuning of search algorithms. 
We used AVM as a search algorithm for the experiment. 
Since AVM does not require any specific parameter tuning, there is no possibility of this threat in our experiment. 
For the comparison with existing constraint solvers, we utilized default recommended parameter settings to mitigate the potential impact of this threat.
	\section{Related Works} \label{rw}
This section provides a comparative analysis of our work with the related literature. 
In the following subsections, first, we relate our work with code-based approaches followed by specification-based approaches, search space reduction, and lastly domain reduction techniques. 
 
\subsection{Code-based Approaches}
In works related to code-based approaches, some approaches target branch coverage~\cite{gupta1998automated,gupta2000generating}, path coverage~\cite{lakhotia2007multi}, and multiple conditions coverage~\cite{ghani2009automatic} for generating test data. 
A few approaches also target MC/DC criterion ~\cite{yu2006comparison,woodward2006relationship,awedikian2009mcdc}. 
In this regard, Godboley \textit{et al.}~\cite{godboley2013increase} proposed an approach to improve MC/DC coverage using code transformations. 
Das and Mall~\cite{das2013automatic} proposed a technique that transforms code under test to achieve MC/DC coverage. 
Li \textit{et al.}~\cite{li2017improving} proposed an approach to improve MC/DC coverage and fault detection effectiveness using combinatorial testing. 
In addition to the previously mentioned approaches, some approaches aim at addressing multiple coverage criteria and incorporate the solutions archiving concept. 
Rojas \textit{et al.}~\cite{rojas2017detailed} performed an empirical study to analyze the coverage of the program under test using the whole test suite generation approach with archiving or covered targets. 
Similarly, Panichella \textit{et al.}~\cite{panichella2018automated} proposed a many-objective search algorithm (DynaMOSA) that dynamically selects coverage targets for test case generation. 
Cegin and Rastocny~\cite{cegin2020test} proposed an approach that uses reinforcement learning to maximize the MC/DC coverage of the code under test. 
In recent work, Godboley \textit{et al.}~\cite{godboley2021toward} presented a method to optimize input programs' MC/DC coverage of input programs. 
The main advantage of our methods over the above-mentioned approaches is that our approach uses UML models and OCL constraints to generate test data and is independent of source code written in any programming language. 
Furthermore, OCL and UML models are used to generate source code~\cite{chauvel2005code, mehmood2013aspect}, in which OCL constraints are translated into conditions in code~\cite{hemmati2018evaluating}. Our strategy can be applied to achieve MC/DC coverage of the source code generated from these models. While this approach can achieve sufficient code coverage, it is important to note that similar coverage cannot be guaranteed, as reported by Hemmati \textit{et al.}~\cite{hemmati2018evaluating}.

\subsection{Specification-based Approaches}
In the work related to specification-based approaches, Ali \textit{et al.}~\cite{ali2013generating,ali2015improving} proposed an approach to generate test data by solving OCL constraints with search. 
In another work, Ali \textit{et al.}~\cite{ali2016generating} proposed an approach targeting boundary value analysis of OCL constraints to improve test data generation. 
In contrast, our work focuses on test data generation for achieving MC/DC coverage of OCL constraints through CBR and range reduction methods. 
Moreover, Weyuker \textit{et al.}~\cite{weyuker1994automatically} presented different strategies for generating test data using Boolean formulas. Offutt \textit{et al.}~\cite{offutt2003generating} proposed an approach to generate test data from state machines while achieving state machine coverage. 
In contrast to these works, our strategy employs OCL constraints to generate test data specifically targeting MC/DC coverage.

In a recent work~\cite{soltana2020practical}, a hybrid test data generation approach (\textcolor{black}{\pledge}) is proposed that utilizes the search and SMT for solving OCL constraints.
The key difference is that our work supports the MC/DC coverage of OCL constraints while generating test data.
Hemmati \textit{et al.}~\cite{hemmati2018evaluating} evaluated the use of MC/DC criterion on guard conditions of the state machine for the generation of coverage-based test data. 
The evaluation results indicated that the search budget, search space, and execution time are the main limitations of solving industrial constraints with search techniques. Motivated by these findings, we introduced an MC/DC test data generation strategy featuring CBR and range reduction methods designed to optimize the search process. 

Numerous constraint solving approaches based on SAT/Alloy~\cite{przigoda2016verifying,przigoda2018automated,przigoda2019four,soeken2011encoding,wotawa2010generating,van2003uml,krieger2010automatic,erata2018alloyinecore,bordbar2005uml2alloy}, satisfiability modulo theories (SMT)~\cite{cantenot2014test, cantenot2013strategies, semerath2013validation, semerath2017formal, semerath2020automated, wu2020verifying, wu2016generating, wu2022query,clavel2010checking}, CSP~\cite{ulke2017partial, gogolla2005validating, gogolla2018achieving, gogolla2020metrics, mokhtari2020validation, ferdjoukh2015instantiation, gonzalez2012emftocsp,aichernig2005test,cabot2008verification,packevicius2013test,cabot2014verification}, partition analysis~\cite{benattou2002generating,li2007test,weissleder2007deriving}, and theorem prover~\cite{brucker2011specification} are also available on in literature. Next, we present these approaches and subsequently relate them to our work. 

\textbf{SAT/Alloy.}
Przigoda \textit{et al.}~\cite{przigoda2016verifying, przigoda2018automated, przigoda2019four} introduced verification of structural and behavioral UML/OCL models in symbolic form using SAT solvers. 
Soeken \textit{et al.}~\cite{soeken2011encoding} demonstrated the encoding of OCL constraints into bit-vector formats and the verification of UML/OCL models using SAT solvers.
Wotawa \textit{et al.}~\cite{wotawa2010generating} proposed an approach that transforms code into constraints, test case generation into a constraint satisfaction problem, and uses MINION~\cite{gent2006minion} to solve generate distinct tests.
Van \textit{et al.}~\cite{van2003uml} introduced a UML-CASTING tool that generates test cases by using UML models and solving OCL constraints.
Krieger \textit{et al.}~\cite{krieger2010automatic} presented an approach that converts operation contracts into Boolean formulas and solves those constraints using the SAT solver.

\textbf{SMT.}
Cantenot \textit{et al.}~\cite{cantenot2013strategies} analyzed various SMT-based constraint-solving strategies for UML/OCL using real-world case studies.
Cantenot \textit{et al.}~\cite{cantenot2014test} proposed a framework that converts UML/OCL models to SMT formats and generates test sequences using animation strategies.
Semerath \textit{et al.}~\cite{semerath2013validation, semerath2017formal} proposed an approach that transforms metamodel constraints in first-order logic to be solved by SMT solver. 
Semerath \textit{et al.}~\cite{semerath2020automated} presented a model generation approach utilizing SMT solver for satisfying constraints. 
Hao Wu proposed various techniques~\cite{wu2016generating, wu2020verifying, wu2022query} that use SMT solvers to satisfy model coverage criteria and verify operational contracts and UML class models.
Clavel \textit{et al.}~\cite{clavel2010checking} proposed a mapping technique for converting the OCL subset into first-order logic and utilizing this to analyze the satisfiability of OCL constraints.

\textbf{CSP.}
Ulke \textit{et al.}~\cite{ulke2017partial} proposed a method for partial evaluation of OCL constraints by formulating them as CSP problems.
Gogolla \textit{et al.}~\cite{gogolla2005validating, gogolla2018achieving} studied UML/OCL models verification and validation with the USE tool. 
Gogolla and Stüber~\cite{gogolla2020metrics} developed a metric for OCL constraints verification based on an expert study.
Ferdjoukh \textit{et al.}~\cite{ferdjoukh2015instantiation} presented a model generation approach that converts metamodel into CSP and uses a CPS solver to check conformance.
Gonzalez \textit{et al.}~\cite{gonzalez2012emftocsp} introduced an EMFtoCSP tool that provides automatic verification of EMF models by transforming them into CPS problems.
Hilken \textit{et al.}~\cite{hilken2015uml} proposed an approach for transforming UML/OCL models into based models and using CSP solvers for verification.
Aichernig \textit{et al.}~\cite{aichernig2005test} proposed an approach that transforms test cases into CSP and performs OCL mutations to generate fault-oriented test cases. 
Cabot \textit{et al.}~\cite{cabot2008verification, cabot2014verification} proposed an approach (\textcolor{black}{\umltocsp}) for UML/OCL model verification using CSP solvers. 
Packevicius \textit{et al.}~\cite{packevicius2013test} presented an approach for generating test data targeting complex data types in model constraints. 
Erata \textit{et al.}~\cite{erata2018alloyinecore} introduced AlloyInEcore which identifies inconsistent models and fills in partial models based on metamodel semantics.
Bordbar and Anastasakis~\cite{bordbar2005uml2alloy} presented the UML2Alloy tool that converts the UML model into Alloy and uses a CSP solver to verify OCL constraints. 

\textbf{Partition Analysis and Theorem Prover.}
Benattou \textit{et al.}~\cite{benattou2002generating} presented a partition analysis method for OCL constraints to generate test data. 
Li \textit{et al.}~\cite{li2007test} proposed an approach to generate test cases from UML sequence diagrams annotated with OCL constraints.  
Wei{\ss}leder and Schlingloff~\cite{weissleder2007deriving} presented an approach to infer boundary values from OCL constraints using input partitioning. 
Babikian \textit{et al.}~\cite{babikian2020automated} proposed a model generation approach that uses theorem provers to solve first-order logic representing metamodel and constraints.
Brucker \textit{et al.}~\cite{brucker2011specification} proposed an approach for test data generation using theorem provers to solve OCL constraints.


In all the aforementioned approaches related to CSP, SAT/Alloy, SMT, and theorem provers, our work distinguishes itself in \textcolor{black}{two} main aspects. First, our work addresses the lack of support for OCL collection operations by these works, which are common in industrial-level constraints. Second, none of the mentioned approaches support test data generation with a focus on achieving MC/DC coverage of OCL constraints, whereas our approach specifically targets this criterion.

\subsection{Search Space Reduction}
Several techniques for search space reduction have been discussed in the literature for CSP and SAT solvers.
Pipatsrisawat \textit{et al.}~\cite{pipatsrisawat2007clone} proposed an approach for lower bound computation and search reduction for MAXSAT solver - an optimized version of SAT. 
Williams \textit{et al.}~\cite{williams2003backdoors} proposed an approach for reducing the combinatorial explosion problem of SAT and CSP solvers which helps in search reduction. 
The idea is to compute a set of variables from the given constraints that makes constraint satisfaction easy. 
Grandcolas \textit{et al.}~\cite{grandcolas2007filtering} proposed several search space pruning strategies to reduce the number of states or actions while SAT solving. 
Charnley \textit{et al.}~\cite{charnley2006automatic} proposed an approach for optimizing constraint solving for CSP solvers and theorem provers. 
The above-mentioned approaches utilize space reduction to facilitate constraint satisfaction for CSP/SAT solvers, whereas our proposed range reduction method focuses on improving search-based test data generation for MC/DC. 
As range reduction involves computing variables' range internally based on constraints, our method contributes to advancing search-based constraint solvers.

\subsection{Domain Reduction Techniques}
Many works have also discussed domain reduction for test data generation. 
Offutt \textit{et al.}~\cite{demilli1991constraint, offutt1999dynamic} proposed domain reduction for the test data generation process. 
Their approach is based on expression substitution, value backtracking, and arbitrary splitting mechanisms that are only suitable for dependent constraints. 
This approach can not apply to MC/DC constraints because these constraints, even if dependent, must be solved independently to generate test data that adheres to MC/DC combinations. Therefore, our range reduction method is designed to work specifically with individual MC/DC constraints. 
McMinn \textit{et al.}~\cite{mcminn2012input} proposed an irrelevant attribute reduction strategy for improving the performance of the automated test data generation process. 
Their approach works for source code whereas our range reduction method specifically addresses OCL constraints. Additionally, eliminating irrelevant attributes is unsuitable for MC/DC constraints, as each attribute in MC/DC constraints holds equal importance for test data generation.

Harman \textit{et al.}~\cite{harman2009automated} proposed a program slicing-based input domain reduction technique to generate test data for aspect-oriented programs.
Their approach operates at the source code level and is specific to aspect-oriented programs, while our methods are designed for handling OCL constraints.
Harman \textit{et al.}~\cite{harman2007impact} presented a theoretical and empirical analysis to investigate the impact of search space reduction on test data generation. 
The results of both types of analysis indicate that reducing search space enhances the performance of search algorithms. 
Arcuri and Yao~\cite{arcuri2008search} introduced search space reduction for Java containers to minimize the computation cost of search algorithms. 
Ribeiro \textit{et al.}~\cite{ribeiro2009test} introduced the input domain reduction strategy to generate tests for object-oriented Java programs. 
Rojas \textit{et al.}~\cite{rojas2016seeding} analyzed different seeding strategies that are used to generate test cases for Java programs. 
The works mentioned above are specifically designed for Java programs, while our methods are focused on reducing the range for MC/DC constraint solving.
Clariso \textit{et al.}~\cite{clariso2017smart} presented a technique that focuses on assisting users in selecting bounds for UML/OCL model verification, whereas our range reduction method automatically computes ranges for MC/DC constraints to facilitate test data generation. 
Bidgoli and Haghighi~\cite{bidgoli2018new} proposed an approach to reduce search space for a program input variable. 
Their approach is confined to individual clauses, whereas our range reduction method considers all independent and dependent clauses used in MC/DC constraints, which is essential for generating test data that adheres to MC/DC combinations. 
Das and Pratihar~\cite{das2019new} proposed a dynamic search space reduction approach for genetic algorithms based on the diversity of the population.  
In comparison, our range reduction method computes ranges for MC/DC constraints at the initial stage and applies these reduced ranges before starting the search process. This flexibility enables utilizing our range reduction method with any search algorithm.

	\section{Conclusion}
	\label{con}
	
	A key concern of the avionics industry is the automated testing of safety-critical systems following international safety standards.
	Model-based testing of such systems requires the automated generation of test data from OCL constraints. 
	The existing OCL-based test data generation approaches target statement coverage of  OCL constraints. 
	International safety standards (e.g., DO-178C) suggest MC/DC criterion for the testing of safety-critical systems. 
	In this paper, we propose an effective way to automate MC/DC test data generation during model-based testing.
	For this purpose, we develop a strategy that utilizes CBR and range reduction heuristics designed to solve MC/DC-tailored OCL constraints. 
	\textcolor{black}{We performed an empirical study to compare our proposed strategy for MC/DC test data generation using original AVM (AVMo), AVM with CBR (AVMc), AVM with range reduction (AVMr), and AVM with the combination of CBR and range reduction (AVMrc). We also empirically compared our proposed CBR and range reduction methods with two constraint solvers, namely UMLtoCSP and PLEDGE.}
	We used 129 OCL constraints from six case studies belonging to different domains with varying sizes and complexity. 
	The experimental results showed that both AVMc and AVMr perform significantly better when compared with the original AVMo in terms of the ability to efficiently solve OCL constraints.
	The results also indicated that the combination of both approaches for solving MC/DC constraints is a viable option to achieve a high success rate in less time.
    \textcolor{black}{Moreover, the results of the comparison with UMLtoCSP and PLEDGE showed that AVMrc outperformed by solving a higher percentage of MC/DC constraints. Nevertheless, in terms of time, UMLtoCSP and PLEDGE indicated better performance.}
	In the future, we plan to conduct an experiment to analyze the effect of the scaling factor used in range reduction for solving MC/DC-tailored OCL constraints.

	\section*{Acknowledgements}
	This research work is supported through a research grant titled \lq Establishment of National Center of Robotics and Automation (NCRA)\rq\space by Higher Education Commission (HEC) Pakistan.

    
    \bibliographystyle{plainnat}
	\bibliography{refs}

\end{document}